 \documentclass[final,3p,times]{elsarticle}

\usepackage{graphicx}

\usepackage{amssymb}
\usepackage{latexsym,graphicx,amssymb,amsmath,mathrsfs}
\usepackage{setspace,bm}
\usepackage{hyperref}

\biboptions{sort&compress}

\graphicspath{{./Figures/}{./real_time_figure/}{./imaginary_time_figures/}}

\newcommand{\diff}{\mathrm{d}}
\newcommand{\p}{\partial}
\newcommand{\ve}{\varepsilon}
\newcommand{\Diff}{{\mathcal{D}}}

\newcommand{\be}{\begin{equation}}      
\newcommand{\ee}{\end{equation}}      
\newcommand{\bea}{\begin{eqnarray}}      
\newcommand{\eea}{\end{eqnarray}}

\newcommand{\ka}{K{\" a}hler }
\newcommand{\schroedinger}{Schr\"odinger }
\newcommand{\im}{\mathrm{i}}

\journal{Annals of Physics}

\begin{document}
\begin{flushright}
RIKEN-QHP-156
\end{flushright}

\begin{frontmatter}


\title{Real-time Feynman path integral with Picard--Lefschetz theory\\ and its applications to quantum tunneling}


\author{Yuya Tanizaki}
\ead{yuya.tanizaki@riken.jp}
\address{Department of Physics, The University of Tokyo, Tokyo 113-0033, Japan}
\address{Theoretical Research Division, Nishina Center, RIKEN, Wako 351-0198, Japan}

\author{Takayuki Koike}
\ead{tkoike@ms.u-tokyo.ac.jp}
\address{Graduate School of Mathematical Sciences, The University of Tokyo, Tokyo 153-8914, Japan}

\begin{abstract}
Picard--Lefschetz theory is applied to path integrals of quantum mechanics, in order to compute real-time  dynamics directly. After discussing basic properties of real-time path integrals on Lefschetz thimbles, we demonstrate its computational method in a concrete way by solving three simple examples of quantum mechanics. 
It is applied to quantum mechanics of a double-well potential, and quantum tunneling is discussed. We identify all of the complex saddle points of the classical action, and their properties are discussed in detail. 
However a big theoretical difficulty turns out to appear in rewriting the original path integral into a sum of path integrals on Lefschetz thimbles. 
We discuss generality of that problem and mention its importance. 
Real-time tunneling processes are shown to be described by those complex saddle points, and thus semi-classical description of real-time quantum tunneling becomes possible on solid ground if we could solve that problem. 
\end{abstract}


\end{frontmatter}


\section{Introduction}\label{sec:intro}
Quantum real-time dynamics has been an important topic for vast areas of physics, especially related to nonequilibrium phenomena. 
Since quantum mechanics deals superposition of probability amplitudes, the Feynman path integral says that time-development of quantum states can be viewed as a sum of amplitudes $\exp\left({\im} S[x(t)]/\hbar\right)$ over all possible spacetime paths $x(t)$~\cite{Feynman:1948ur, FeynmanHibbs196506}. 
However, such summations show truly bad convergence due to the highly oscillatory factor without any suppressions. 
This leads to non-existence of the real-time path integral as the standard integration theory~\cite{cameron1960family}, and also makes difficult its numerical simulation due to the severe sign problem. 
From measure theoretical point of view, its close relation to the Wiener integration is first revealed by Kac when time $t$ is replaced by the imaginary time $-\im \tau$~\cite{kac1951} (see also Refs.~\cite{nelson1964,simon1979functional}). 
Imaginary-time path integral provides a convenient formalism also for numerical computations for thermal equilibrium systems, but analytic continuation must be performed to obtain real-time dynamics. 
Realization of real-time path integrals based on the measure theory itself is an interesting task in mathematical physics, but it will also provide a convenient framework for direct simulations of quantum real-time phenomena. 

Recently, Witten proposed an application of Picard--Lefschetz theory to Feynman path integral~\cite{Witten:2010cx, Witten:2010zr}.  Picard--Lefschetz theory, which is a complex analogue of Morse theory, tells us all possible deformations of an integration contour in the complexified space of an original integration cycle. 
Therefore, we are free from oscillatory integrals by choosing ``nice'' integration cycles, which are called Lefschetz thimbles. 
This technique has a potential to reveal nonperturbative aspects of quantum field theory. Indeed, there is already an interesting suggestion that existence of Lefschetz thimbles around nonperturbative critical points may be closely related to ambiguities of Borel summation of perturbation theory~\cite{Unsal:2012zj, Basar:2013eka, Cherman:2014ofa}. 
Its practical applications are now also being studied especially for solving the sign problem in Monte Carlo simulations of statistical quantum systems; finite--density quantum chromodynamics~\cite{Cristoforetti:2012su, Cristoforetti:2013wha, Cristoforetti:2014gsa, Aarts:2013fpa, Fujii:2013sra}, and repulsive Hubbard model in condensed matter physics~\cite{Mukherjee:2014hsa}. 

In this paper, Picard--Lefschetz theory is applied to real-time path integrals of quantum systems. 
After reviewing the path-integral formalism based on Picard--Lefschetz theory, we first study fundamental properties of path integrals on Lefschetz thimbles. 
In order to show how this method works, simple examples of quantum mechanics are considered, and Feynman kernels are computed for free particles and for a harmonic oscillator by using path integrals on Lefschetz thimbles. 
We will be able to see at least for these examples that real-time path integrals are now well-defined based on the standard integration theory on Lefschetz thimbles. 

As an application of this formalism, quantum mechanics of a double-well potential is considered, and quantum tunneling is discussed. We identify all saddle points of the classical action in the complexified space of spacetime paths, and their properties are discussed in detail. 
However a big theoretical difficulty turns out to appear in rewriting the original path integral into a sum of path integrals on Lefschetz thimbles. 
We discuss generality of that problem and mention its importance for future numerical computations of path integrals with sign problem. 
We do not solve this problem in a direct way, but we argue that real-time tunneling must be described by highly-oscillatory complex classical solutions, by scrutinizing those solutions both in real-time and imaginary-time formalisms. 
Therefore, we can obtain exact semi-classical description of real-time quantum tunneling on solid ground if we could solve that theoretical difficulty. 

This paper is organized as follows. 
In Section~\ref{sec:basic}, we first briefly review applications of Picard--Lefschetz theory to path integrals. After that, we discuss basic properties of path integrals on each Lefschetz thimble. 
In Section~\ref{sec:examples}, we demonstrate how to compute real-time path integrals of quantum mechanics by applying this formalism. 
Real-time path integrals are calculated for a free particle on a line in Section~\ref{sec:free}, a free particle on a circle in Section~\ref{sec:free_circle}, and a harmonic oscillator in Section~\ref{sec:ho}.  
Through these simple examples, basic strategies for application of Picard--Lefschetz theory will be established for taking into account topological effects and Maslov--Morse index of closed trajectories. 
In Section~\ref{sec:tunneling}, tunneling phenomena are studied through our formalism of real-time path integral. 
After identifying all the complex classical solutions of the double-well potential in Section~\ref{sec:label_dw_solutions}, we discuss various aspects of those classical solutions in Section~\ref{sec:discussion_tunneling} in order to find out their relation to quantum tunneling. 
In Section~\ref{sec:summary}, we summarize our result. 
In \ref{sec:form}, we give a concise review of  Picard--Lefschetz theory for an application to oscillatory integrals. 

\section{Basic properties of Picard--Lefschetz theory for path integral}\label{sec:basic}

In this section, we first review the way to apply Picard--Lefschetz theory to path integrals. 
In this formalism, infinite dimensional oscillatory integrals become summations of well-defined integrals on appropriate integration cycles, called Lefschetz thimbles. 
We show that a quantum equation of motion holds on each Lefschetz thimble. 
Possible configurations of Lefschetz thimbles are studied by studying properties of Morse's downward flow equations for real-time path integrals. 

\subsection{Application of the Picard--Lefschetz theory to path integrals}
We consider quantization of a classical system described by an action functional $\mathcal{I}[x]=\im S[x(t)]$, where the imaginary unit $\im=\sqrt{-1}$ is multiplied to the classical action $S[x(t)]=\int \diff t L(x,\diff x/\diff t)$ for later convenience. 
Path-integral quantization says that the transition amplitude from $x_i$ to $x_f$ after time $(t_f-t_i)$ is described by summation of amplitudes $\exp\mathcal{I}[x]/\hbar$ over the space of all possible paths $\mathcal{Y}=\{x:[t_i,t_f]\to \mathbb{R}|x(t_i)=x_i, x(t_f)=x_f\}$, 
\be
K(x_f,t_f;x_i,t_i)=\int_{\mathcal{Y}}\Diff x \exp{\mathcal{I}[x]\over \hbar}. 
\label{eq:feynman}
\ee
However, since the action $\mathcal{I}[x]$ is purely imaginary for any $x:[t_i,t_f]\to \mathbb{R}$, this expression (\ref{eq:feynman}) must be understood as a limit of oscillatory integrals. 

Application of Picard--Lefschetz theory to oscillatory integrals in (\ref{eq:feynman}) is briefly given without any proof in the following. For more detailed explanations, see \cite{Witten:2010cx,Witten:2010zr} and also \ref{sec:form}. 
\begin{enumerate}
\item Find all solutions of the classical equation of motion, $\delta \mathcal{I}=0$, in the complexified space of paths $\mathcal{X}=\{z:[t_i,t_f]\to \mathbb{C}|z(t_i)=x_i, z(t_f)=x_f\}$. Those solutions are denoted by $z_{\sigma}$ with a label $\sigma\in\Sigma$. 
\item Calculate the Lefschetz thimble $\mathcal{J}_{\sigma}$ associated to each classical stationary path $z_{\sigma}$ by solving the downward flow equation
\be 
{\p z(t;u)\over \p u}=-\overline{{\delta \mathcal{I}[z(t;u)]\over \delta z(t;u)}}, 
\label{eq:basic02}
\ee 
with the boundary conditions $z(t;-\infty)=z_{\sigma}(t)$, and $z(t_i,u)=x_i$ and $z(t_f,u)=x_f$. 
Since $\mathrm{Re}\; \mathcal{I}\to -\infty$ and $\mathrm{Im}\; \mathcal{I}$ is a constant along downward flows, the path integral on each Lefschetz thimble $\mathcal{J}_{\sigma}$ shows good convergence. 
\item With some integers $n_{\sigma}$, the original path integral (\ref{eq:feynman}) can now be rewritten as 
\be
K(x_f,t_f;x_i,t_i)=\sum_{\sigma\in\Sigma}n_{\sigma} \int_{\mathcal{J}_{\sigma}}\Diff z \exp{\mathcal{I}[z]\over \hbar}. 
\ee
Those integral coefficients $n_{\sigma}$ are defined by the number of intersecting points between the space of real paths $\mathcal{Y}$ and upward flows $\mathcal{K}_{\sigma}$ from $z_{\sigma}$ in the homological sense: $n_{\sigma}=\langle \mathcal{Y},\mathcal{K}_{\sigma}\rangle$. 
\end{enumerate}
In the above procedure, we implicitly assume that Lefschetz thimbles are well-defined. This assumption breaks down if a downward flow (\ref{eq:basic02}) connects two distinct critical points. However, since $\mathrm{Im}\; \mathcal{I}$ is a conserved quantity, this does not happen except for special parameters or symmetries in $\mathcal{I}$. 
In this section, we consider generic cases and assume that all the Lefschetz thimbles are well-defined. 

\subsection{Quantum equation of motions on Lefschetz thimbles}
We consider equation of motions at the quantum level. In the language of path integral, quantum equation of motions can be denoted as 
\be
\int \Diff x {\delta \mathcal{I}[x]\over \delta x(t)}\exp {\mathcal{I}[x]\over \hbar}=0 
\label{eq:eqm01}
\ee
for any boundary conditions at $t=t_i$ and $t_f$. In this part, we would like to show that this relation (\ref{eq:eqm01}) holds for each Lefschetz thimble independently, that is, for any $\sigma\in\Sigma$ 
\be
\int_{\mathcal{J}_{\sigma}}\Diff z {\delta \mathcal{I}[z]\over \delta z(t)}\exp{\mathcal{I}[z]\over \hbar}=0. 
\label{eq:eqm02}
\ee

For that purpose, we consider finite-dimensional analogue of (\ref{eq:eqm02}). Let $\mathcal{I}$ be a holomorphic function of $z=(z^1,\ldots,z^n)$, which has critical points $z_{\sigma}$. Assume that the Lefschetz thimble $\mathcal{J}_{\sigma}$ is well defined for each critical point $z=z_{\sigma}$ of $\mathcal{I}$, then, for $1\le i\le n$, 
\be
\int_{\mathcal{J}_{\sigma}}\diff z^1\cdots \diff z^n {\p \mathcal{I}(z)\over \p z^i} \exp{\mathcal{I}(z)\over \hbar}=0. 
\label{eq:qem03}
\ee
In order to prove (\ref{eq:qem03}), we prepare following notations: let $\ve=(\ve^1,\ldots,\ve^n)\in\mathbb{C}^n$, then 
\bea
\mathcal{I}_{\ve}(z):=\mathcal{I}(z+\ve), \quad
Z_{\sigma,\ve}:=\int_{\mathcal{J}_{\sigma}} \Diff z \exp{\mathcal{I}_{\ve}(z)\over \hbar}. 
\eea
For proving (\ref{eq:qem03}), it suffices to show that $Z_{\sigma,\ve}=Z_{\sigma,0}$ for $|\ve|\ll 1$, and that 
\be
\left.{\p Z_{\sigma,\ve}\over \p \ve^i}\right|_{\ve=0}=\int_{\mathcal{J}_{\sigma}}\diff^n z{\p \mathcal{I}(z)\over \p z^i}\exp{\mathcal{I}(z)\over \hbar}. 
\label{eq:qem06}
\ee

Clearly, all the critical points of $\mathcal{I}_{\ve}$ can be labeled by the same set of critical points $\Sigma$ of $\mathcal{I}$, and each critical point $z_{\sigma,\ve}$ of $\mathcal{I}_{\ve}$ is given by 
\be
z_{\sigma,\ve}=z_{\sigma}-\ve. 
\label{eq:qem07}
\ee
We can also readily find that the Lefschetz thimble $\mathcal{J}_{\sigma,\ve}$ of $\mathcal{I}_{\ve}$ associated to $z_{\sigma,\ve}$ is given by 
\be
\mathcal{J}_{\sigma,\ve}=\mathcal{J}_{\sigma}-\ve:=\left\{z-\ve| z\in \mathcal{J}_{\sigma}\right\} 
\label{eq:qem08}
\ee
as a subset of $\mathbb{C}^n$. 
When $\ve=0$, $\mathcal{J}_{\sigma, \ve}$ does not intersect $\mathcal{K}_\tau$ if $\sigma\not=\tau$ and it intersects $\mathcal{K}_\sigma$ transversally only at one point, which is an open condition for $\ve$. 
Thus, for sufficiently small parameters $\ve$, 
\be
\langle \mathcal{J}_{\sigma,\ve}, \mathcal{K}_{\tau}\rangle=\delta_{\sigma\tau}, 
\label{eq:qem09}
\ee
which implies that $\mathcal{J}_{\sigma,\ve}=\mathcal{J}_{\sigma}$ as integration cycles. This completes the proof of $Z_{\sigma,\ve}=Z_{\sigma,0}$, and we find that $\p Z_{\sigma,\ve}/\p \ve^i=0$. 

Let us give a proof of (\ref{eq:qem06}) under the following assumption: the action functional $\mathcal{I}:\mathbb{C}^N\to\mathbb{C}$ takes the form 
\be
\mathcal{I}(z)=\sum_{i\not=j}J_{ij}z^i z^j + \sum_{i=1}^{N} V(z^i), 
\label{eq:qem10}
\ee
where $V(z^i)$ is a polynomial of degree $L>2$. Under this assumption (\ref{eq:qem10}), we can find asymptotic behaviors of $\mathcal{I}$ in the limit $|z|\to\infty$ along Lefschetz thimbles $\mathcal{J}_{\sigma}$: putting $V(z^i)=a (z^i)^L+\mathcal{O}((z^i)^{L-1})$, 
\be
\mathcal{I}(z)\sim -|a| \sum_{i=1}^{N}|z^i|^L+\mathcal{O}(|z|^{L-1}). 
\label{eq:qem11}
\ee

Let us estimate the following derivative; 
\bea
\left|{\p \over \p \ve^i}e^{\mathcal{I}_{\ve}(z)-\mathcal{I}(z)}\right|=
\left|{\p \mathcal{I}_{\ve}(z) \over \p \ve^i}e^{\mathcal{I}_{\ve}(z)-\mathcal{I}(z)}\right|
\le \left|\mathcal{P}_{L-1}(z)\right|\exp |\mathcal{P}_{L-1}(z)|. 
\label{eq:qem12}
\eea
In (\ref{eq:qem12}), $\mathcal{P}_M (z)$ refers some polynomials of order $M$, which are independent of $\ve$. 
The right hand side of (\ref{eq:qem12}) is integrable on Lefschetz thimbles $\mathcal{J}_{\sigma}$ with the measure $\exp \mathcal{I}(z)\diff^n z$ thanks to an estimate of asymptotic behaviors (\ref{eq:qem10}). Therefore, Lebesgue's dominated convergence theorem ensures (\ref{eq:qem06}). 

This completes the proof of quantum equations of motion (\ref{eq:qem03}), because we showed that $\p Z_{\sigma,\ve}/\p \ve^i=0$ and (\ref{eq:qem06}). 

\subsection{Constraints on $\Sigma$ and Lefschetz thimbles for real-time path integrals}
Let $\mathcal{I}(z)=\im S(z)$ be an action functional, satisfying $\overline{S(z)}=S(\overline{z})$. This property is equivalent to $\overline{\mathcal{I}(z)}=-\mathcal{I}(\overline{z})$, and then 
\bea
\mathrm{Re}\;\mathcal{I}(z)&=&{\mathcal{I}(z)-\mathcal{I}(\overline{z})\over 2}=-\mathrm{Re}\;\mathcal{I}(\overline{z}), 
\label{eq:constraint01}\\
\mathrm{Im}\;\mathcal{I}(z)&=&{\mathcal{I}(z)+\mathcal{I}(\overline{z})\over 2\im}=\mathrm{Im}\;\mathcal{I}(\overline{z}). 
\label{eq:constraint02}
\eea

For a complex classical solution $z_{\sigma}$ of $\mathcal{I}$, we denote its complex conjugate by $z_{\overline{\sigma}}$: $z_{\overline{\sigma}}=\overline{z_{\sigma}}$. 
Then $\overline{\sigma}\in \Sigma$, that is, $z_{\overline{\sigma}}$ is also a critical point: 
\be
{\p \mathcal{I}\over \p z}(z_{\overline{\sigma}})=-\overline{{\p \mathcal{I}\over \p z}(z_{\sigma})}=0. 
\label{eq:constraint03}
\ee
$\mathcal{J}_{\overline{\sigma}}$ and $\mathcal{K}_{\overline{\sigma}}$ are complex conjugates of $\mathcal{K}_{\sigma}$ and $\mathcal{J}_{\sigma}$, respectively. 
Indeed, an upward flow equation is related to a downward flow equation by complex conjugation: 
\be
{\p z\over \p u}=\overline{-{\p \mathcal{I}(z)\over \p z}}\; \Leftrightarrow \; 
{\p \overline{z}\over \p u}=-{\p \mathcal{I}({z})\over \p {z}}=\overline{\p \mathcal{I}(\overline{z})\over \p \overline{z}}. 
\label{eq:constraint04}
\ee
Therefore, 
\be
\mathcal{J}_{\overline{\sigma}}=\overline{\mathcal{K}_{\sigma}}:=\{z|\overline{z}\in \mathcal{K}_{\sigma}\},\; \mathcal{K}_{\overline{\sigma}}=\overline{\mathcal{J}_{\sigma}}. 
\label{eq:constraint05}
\ee

\begin{figure}[t]
\centering
\includegraphics[scale=0.6]{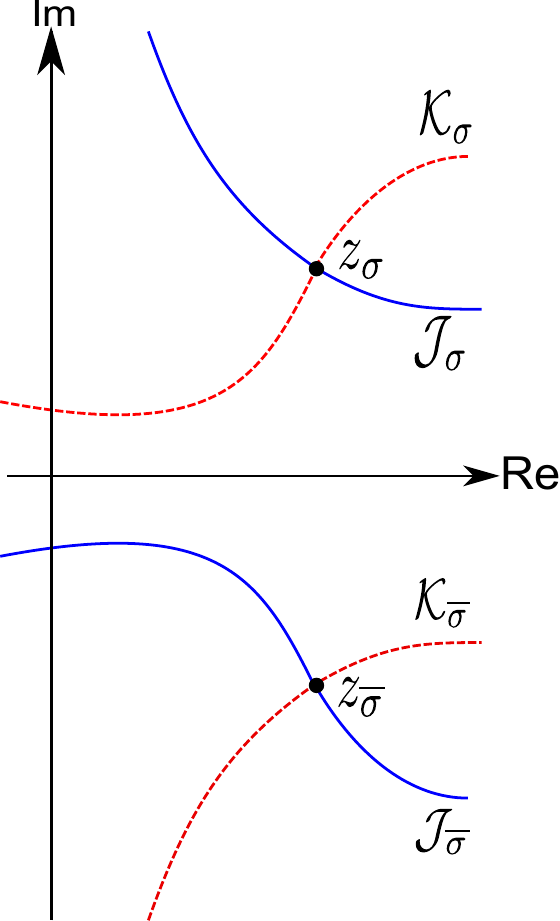}(a)\qquad
\includegraphics[scale=0.6]{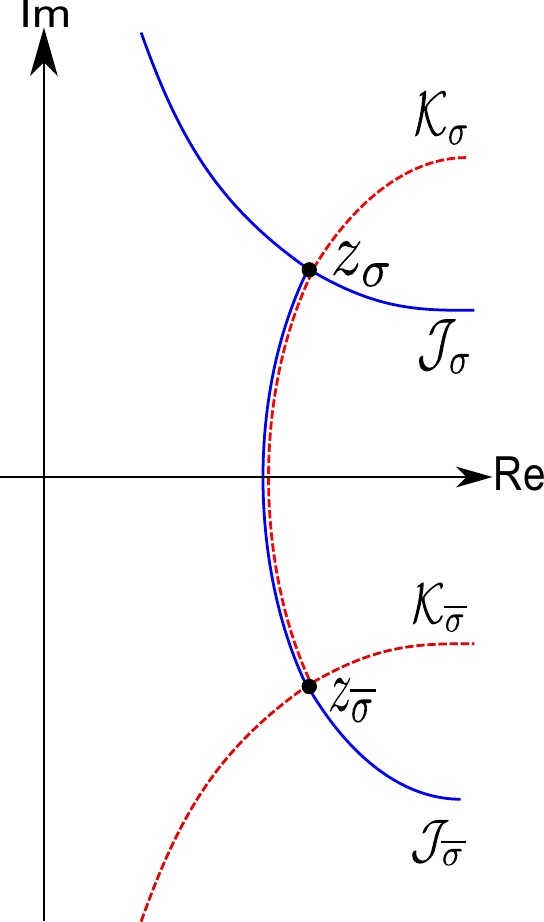}(b)\qquad
\includegraphics[scale=0.6]{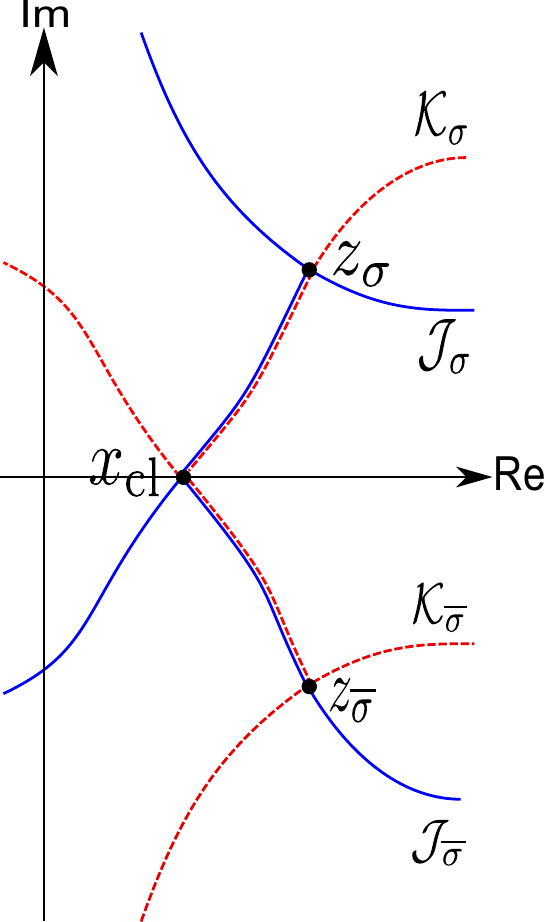}(c)
\caption{Possible behaviors of Lefschetz thimbles for non-real classical solutions in real-time path integrals. Red dashed lines show upward flows, and blue solid ones show downward flows. (a)~Upward flows from $z_{\sigma}$ do not intersect $\mathcal{Y}$. (b)~$z_{\sigma}$ and its complex conjugate $z_{\overline{\sigma}}$ are connected by an upward/downward flow. (c)~There exists a real classical solution $x_{\mathrm{cl}}$ with $\mathrm{Im}\;\mathcal{I}[x_{\mathrm{cl}}]=\mathrm{Im}\;\mathcal{I}[z_{\sigma}]$, and those critical points are connected by downward/upward flows. }
\label{fig:constraint}
\end{figure}

Due to (\ref{eq:constraint01}), we can generally assume that 
\be
\mathrm{Re}\; \mathcal{I}(z_{\sigma})<0,\; \mathrm{Re}\; \mathcal{I}(z_{\overline{\sigma}})>0. 
\label{eq:constraint06}
\ee
This ensures that $\langle \mathcal{Y},\mathcal{K}_{\overline{\sigma}}\rangle=0$ since $\mathrm{Re}\; \mathcal{I}|_{\mathcal{Y}}=0$, and the integration on $\mathcal{J}_{\overline{\sigma}}$ does not contribute to the real-time path integral (\ref{eq:feynman}). 
Several possible behaviors of downward/upward flows with these constraints are shown in Fig.~\ref{fig:constraint}. 
According to this figure, $\mathcal{K}_{\sigma}$ transversally intersects $\mathcal{Y}$ only when an upward flow from $z_{\sigma}$ reach another critical point in the limit $u\to \infty$. 
Since $\mathrm{Im}\;\mathcal{I}$ is conserved along upward/downward flows, those critical points have the same imaginary part of the action functional $\mathcal{I}$. 
Due to (\ref{eq:constraint02}), there is a possibility that $z_{\sigma}$ and its complex conjugate $z_{\overline{\sigma}}$ are connected by the flow equation (\ref{eq:basic02}) (see Fig.~\ref{fig:constraint}(b)). 
If there exists another classical solution $z_{\tau}$ with $\mathrm{Im}\;\mathcal{I}[z_{\sigma}]=\mathrm{Im}\;\mathcal{I}[z_{\tau}]$, $z_{\sigma}$ and $z_{\tau}$ could also be connected by the flow equation (\ref{eq:basic02}). 
In Fig.~\ref{fig:constraint}(c), we show a possible behavior of Lefschetz thimbles if there exists a real classical solution satisfying $\mathrm{Im}\;\mathcal{I}[x_{\mathrm{cl}}]=\mathrm{Im}\;\mathcal{I}[z_{\sigma}]$. 

When there exists a complex classical solution with nonzero $n_{\sigma}$, the Stokes phenomenon necessarily occurs in real-time path integrals. 
In order to make Lefschetz thimbles $\mathcal{J}_{\sigma}$ well-defined as integration cycles, we must replace $\mathcal{I}$ by $e^{+\im 0^+}\mathcal{I}$, or equivalently $\hbar$ by $e^{-\im 0^+}\hbar$.

\subsection{Flow equations around complex classical solutions}

Let us consider the local structure of downward/upward flow equations around a complex classical solutions $z_{\sigma}$. 
We consider a general action $\mathcal{I}=\im \int\diff t\left[{1\over 2}\dot{z}^2-V(z)\right]$, then the downward flow equation (\ref{eq:basic02}) is given by 
\be
{\p z(t;u)\over \p u}=-\im \left({\p^2 \overline{z}(t;u)\over \p t^2}+V'(\overline{z}(t;u))\right). 
\label{eq:semiclassical01}
\ee
Its linearization around the classical solution $z_{\sigma}$ gives 
\be
{\p\over \p u}\Delta z(t;u)=-\im\left({\p^2\over \p t^2}+V''(\overline{z_{\sigma}}(t))\right)\overline{\Delta z}(t;u). 
\label{eq:semiclassical02}
\ee
Let us attempt to find its solution by separating variables: by putting an Ansatz $\Delta z(t;u)=e^{\im\pi/4}\exp(\lambda u)f(t)$ with a positive real number $\lambda$, (\ref{eq:semiclassical02}) becomes an eigenvalue equation 
\be
-\left({\p^2\over \p t^2}+V''(\overline{z_{\sigma}}(t))\right)\overline{f}(t)=\lambda f(t). 
\label{eq:semiclassical03}
\ee
Let us denote $\Omega(t)^2=\mathrm{Re}\; V''(\overline{z_{\sigma}}(t))$ and $\Gamma(t)^2=\mathrm{Im}\; V''(\overline{z_{\sigma}}(t))$, then 
\be
\left(\begin{array}{cc}-\left({\p_t^2}+\Omega(t)^2\right)&-\Gamma(t)^2\\
-\Gamma(t)^2&\left({\p_t^2}+\Omega(t)^2\right)\end{array}\right)
\left(\begin{array}{c} f_1\\ f_2\end{array}\right)
=\lambda \left(\begin{array}{c} f_1\\ f_2\end{array}\right), 
\label{eq:semiclassical04}
\ee
with $f(t)=f_1(t)+\im f_2(t)$. 
Since the $2\times 2$ matrix-valued differential operator on the left hand side of (\ref{eq:semiclassical04}) is self-adjoint on the space of smooth functions with Dirichlet boundary condition, its eigenvalues are real and eigenfunctions associated with different eigenvalues are orthogonal to each other. 
The self-adjoint operator is denoted by $L$: $L=-(\p_t^2+\Omega(t)^2)\sigma_3-\Gamma(t)^2\sigma_1$ using Pauli matrices $\sigma_i$. Let $\ve=\im\sigma_2$, then $\ve^{\dagger}L\ve=-L$ and $\ve\ve^{\dagger}=1$. Therefore, nonzero eigenvalues of $L$ must be paired with its counterpart of opposite sign. 
This property is quite important for the orthonormal property of eigenfunctions in (\ref{eq:semiclassical03}). 

Let $\lambda_a$ and $\lambda_b$ be eigenvalues of $L$ associating eigenfunctions $(f_{a,1},f_{a,2})$ and $(f_{b,1},f_{b,2})$, respectively, and assume that $|\lambda_a|\not=|\lambda_b|$. Since $(f_{b,2},-f_{b,1})=\ve^{\dagger}(f_{b,1},f_{b,2})$ is also an eigenfunction with the eigenvalue $-\lambda_b$, orthogonality shows that 
\be
\int\diff t(f_{a,1}f_{b,1}+f_{a,2}f_{b,2})=\int\diff t(f_{a,1}f_{b,2}-f_{a,2}f_{b,1})=0. 
\label{eq:semiclassical05}
\ee
Putting $f_{a}=f_{a,1}+\im f_{a,2}$ and $f_{b}=f_{b,1}+\im f_{b,2}$, (\ref{eq:semiclassical05}) can be written as 
\be
\int \diff t \overline{f_a}f_b=0, 
\ee
which is nothing but the orthogonality for (\ref{eq:semiclassical03}). 
Eigenfunctions with positive eigenvalues span the tangent space of $\mathcal{J}_\sigma$ at $z_{\sigma}$, and those with negative eigenvalues span that of $\mathcal{K}_\sigma$. Those tangent spaces are related by multiplication of $\ve$, and we can explicitly see how $\mathcal{J}_\sigma$ and $\mathcal{K}_\sigma$ intersect each other. 

Let $\{\lambda_n\}_{n=1}^{\infty}$ be the set of positive eigenvalues of $L$, and $\{f_n\}_n$ be corresponding normalized eigenfunctions, then the tangent space of the Lefschetz thimble $\mathcal{J}_{\sigma}$ at $z_{\sigma}$ is given by 
\be
\left\{\left.\sum_{n=1}^{\infty}a_n e^{\im\pi/4}f_n(t)\; \right| \; a_n\in\mathbb{R}\right\}. 
\ee
For small variations $\Delta z(t)=\sum_n a_n e^{\im\pi/4}f_n(t)$ around $z_{\sigma}$, the classical action $\mathcal{I}$ behaves up to $\mathcal{O}(\Delta z^2)$ as 
\bea
\mathcal{I}[z_{\sigma}+\Delta z]&=&\mathcal{I}[z_{\sigma}]+\sum_{n,m}{a_n a_m\over 2}\int \diff t f_n(t)\left({\p^2\over \p t^2}+V''(z_{\sigma}(t))\right)f_m(t)\nonumber\\
&=&\mathcal{I}[z_{\sigma}]-\sum_{n=1}^{\infty}{\lambda_n\over 2}a_n^2. 
\eea
Therefore, at least for small quantum fluctuations around a critical point $z_{\sigma}$, the real-time path integral on the Lefschetz thimble $\mathcal{J}_\sigma$ can be understood as a Wiener integration. 
We expect that its integration measure would become well-defined based on the standard integration theory also for large fluctuations.

\section{Some simple examples in quantum mechanics}\label{sec:examples}
In this section, we demonstrate applications of Picard--Lefschetz theory to real-time path integrals through three simple examples of quantum mechanics; free particles on a line and on a circle, and a harmonic oscillator. 
Since these three systems can be solved exactly in a simple way, it is quite instructive to learn how Picard--Lefschetz theory works for real-time path integrals. 
Furthermore, the real--time path integral turns out to be constructed based on Lebesgue integration theory on Lefschetz thimbles. 

\subsection{Free particle on a line}\label{sec:free}
In this subsection, we consider path integral for a free particle in order to demonstrate how Picard--Lefschetz theory works, and also to provide new mathematical rigorous definition of real-time path integral. 
That is, the fundamental solution $K_{\mathrm{free}}(x_f,t_f:x_i,t_i)$ of the \schroedinger equation, 
\be
i\hbar{\p\over \p t}\psi(x,t)=-{\hbar^2\over 2}{\p^2\over \p x^2}\psi(x,t), 
\label{eq:free_schr}
\ee
will be constructed by applying the Picard--Lefschetz theory to the path integral (\ref{eq:feynman}). 

In this system, the classical action in~(\ref{eq:feynman}) is given by,
\be
\mathcal{I}[z]={\im} \int \diff t\; {1\over 2}\left({\diff z\over \diff t}\right)^2,
\label{eq:free01}
\ee
which must be regarded as a holomorphic functional on $\mathcal{X}=\{z:[t_i,t_f]\to \mathbb{C}| z(t_i)=x_i, z(t_f)=x_f\}$. 
Critical points $z_{\mathrm{cl}}$ of this functional are determined by the Euler--Lagrange equation: 
\be
{\diff^2 z_{\mathrm{cl}}(t)\over \diff t^2}=0. 
\label{eq:free02}
\ee
Combining this with the boundary condition $z_{\mathrm{cl}}(t_i)=x_i$ and $z_{\mathrm{cl}}(t_f)=x_f$, we find the solution 
\be
z_{\mathrm{cl}}(t)=(x_f-x_i) {t-t_i\over t_f-t_i}+x_i, 
\label{eq:free03}
\ee
which gives $\mathcal{I}[z_{\mathrm{cl}}]=\im{(x_f-x_i)^2/2(t_f-t_i)}$. 
It is important to notice that this solution~(\ref{eq:free03}) is a real function, which is in the domain of original integration cycle in~(\ref{eq:feynman}). 

We introduce a \ka metric 
\be
\diff s^2=\int \diff t\; {1\over 2}(\delta z(t)\otimes \delta\overline{z}(t)+\delta \overline{z}(t)\otimes \delta z(t)), 
\label{eq:metric}
\ee
on the space of complexified field configurations, in which $\delta z(t)$ and $\delta \overline{z}(t)$ are regarded as a basis of the cotangent space. 
This associates the \ka form $\omega=\int \diff t{\im \over 2}\delta z(t)\wedge \delta \overline{z}(t)$. 
The Lefschetz thimble $\mathcal{J}$ associates the complex classical path $z_{\mathrm{cl}}(t)$, and it is defined as a set of fields connected to the classical solution via downward flows: 
\be
\mathcal{J}=\left\{z(t;0)\in\mathcal{X}\left|z(t;-\infty)=z_{\mathrm{cl}}(t),\; {\p z(t;u)\over \p u}=-\overline{\left({\delta \mathcal{I}[z(t;u)]\over \delta z(t;u)}\right)}\right.\right\}. 
\label{eq:free05}
\ee
This is a middle-dimensional cycle in the space of complexified paths. 
Now the functional integration (\ref{eq:feynman}) for the Feynman kernel of the free particle $K_{\mathrm{free}}$ can be rewritten as 
\be
K_{\mathrm{free}}(x_f,t_f;x_i,t_i)=\int_{\mathcal{J}}\Diff z \exp{\mathcal{I}[z] \over \hbar}.  
\label{eq:free06}
\ee
Since the critical point $z_{\mathrm{cl}}$ is given as real functions of $t$, we can immediately use the formula (\ref{app:PicardLefschetz}) to derive this result. 

Let us calculate the Lefschetz thimble $\mathcal{J}$ explicitly. Denote $z(t;u)=z_{\mathrm{cl}}(t)+\Delta z(t;u)$, where $\Delta z\to 0$ as $u\to -\infty$.  The analytic form of the downward flow equation in (\ref{eq:free05}) is given as a set of parabolic partial differential equations, 
\be
{\p \over \p u}\Delta z (t;u)=-{\im}{\p^2\over \p t^2}\overline{\Delta z(t;u)}, 
\label{eq:free07}
\ee
under the boundary condition $\Delta z(t;-\infty)=0$, and $\Delta z(t_i;u)=\Delta z(t_f;u)=0$.   
The set of solutions $\Delta z(t;u)$ for (\ref{eq:free07}) are spanned by 
\be
\Delta z_{\ell}(t;u)=e^{\im \pi/4}\exp\left({\pi^2\ell^2\over 4 (t_f-t_i)^2}u\right)\sin \left(\pi \ell {t-t_i\over t_f-t_i}\right),
\label{eq:free10}
\ee
with $\ell \in \mathbb{Z}_{>0}$. Therefore, an element of Lefschetz thimbles $\mathcal{J}$ can be described as 
\be
z(t)=z_{\mathrm{cl}}(t)+e^{\im \pi/4}\sum_{\ell=1}^{\infty}a_\ell \sin\left(\pi\ell{t-t_i\over t_f-t_i}\right)
\label{eq:free11}
\ee
with $(a_\ell)\in \ell^2(\mathbb{R})$. 

The Jacobian in the change of functional integral ``measure'' is $+1$. To see this, we must notice that the induced metric on the Lefschetz thimble $\mathcal{J}$ from (\ref{eq:metric}) is given by 
\be
\left.\diff s^2\right|_{\mathcal{J}}=\sum_{\ell,\ell'\not=0}\diff a_\ell \diff a_{\ell'}\int_{t_i}^{t_f}\diff t \sin\left(\pi\ell{t-t_i\over t_f-t_i}\right)\sin\left(\pi\ell'{t-t_i\over t_f-t_i}\right)=(t_f-t_i)\sum_{\ell\not=0}(\diff a_\ell)^2. 
\label{eq:free12}
\ee
Therefore, functional integration measure on the Lefschetz thimble $\mathcal{J}$ is proportional to 
\be
\int_{\mathcal{J}}\Diff z= \mathcal{N} \int\prod_{\ell\not=0}\sqrt{\im}\diff a_\ell,  
\label{eq:free13}
\ee
where the normalization factor $\mathcal{N}$ is introduced. Therefore, real-time path integral turns out to be defined as the Wiener integration on Lefschetz thimbles, and its result on each Lefschetz thimble $\mathcal{J}$ becomes 
\be
\int_{\mathcal{J}}\Diff z \exp\mathcal{I}[z]= \mathcal{N}\prod_{\ell\not=0}\left[{\pi \im \over \pi^2\ell^2/2 (t_f-t_i)^2}\right]^{1/2} \exp{\mathcal{I}[z_{\mathrm{cl}}]\over \hbar}. 
\label{eq:free14}
\ee
Here we do not give detailed calculations for the normalization factor $\mathcal{N}$, but it can also be calculated with a more careful treatment of the Wiener integration: 
\be
\int_{\mathcal{J}}\Diff z \exp\mathcal{I}[z]= \sqrt{1\over 2\pi \im \hbar(t_f-t_i)}\exp{\mathcal{I}[z_{\mathrm{cl}}]\over \hbar}. 
\label{eq:free15}
\ee

Let us comment on relationship between real-time and imaginary-time path integrals before closing this section. For that purpose, we formally replace the time $t$ by $-\im e^{\im \varphi}t_{\varphi}$ with $0\le \varphi\le {\pi\over 2}$ in (\ref{eq:free01}) and regard a path as a map $x:[t_i,t_f]\to\mathbb{R},\; t_{\varphi}\mapsto x(t_{\varphi})$. This procedure is called Wick rotation \cite{PhysRev.96.1124}. At $\varphi={\pi\over 2}$, $t_{\varphi}$ is equal to the real time $t$.  The Feynman path integral (\ref{eq:feynman}) becomes 
\be
K_{\mathrm{free}}(x_f,-\im e^{\im \varphi}t_f;x_i,-\im e^{\im\varphi}t_i)=\int\Diff x \exp \left[-{ e^{-\im\varphi}\over \hbar}\int \diff t_{\varphi}{1\over 2}\left({\diff x\over \diff t_{\varphi}}\right)^2\right]. 
\label{eq:wick01}
\ee
When $\varphi=0$, this path-integral measure can be constructed on the space of real paths, and this functional integration is mathematically meaningful from the first~\cite{kac1951, nelson1964, simon1979functional}. 
Let us apply Picard--Lefschetz theory to (\ref{eq:wick01}) for general $\varphi$, so we first define 
\be
\mathcal{I}_{\varphi}[z(t_{\varphi})]=-{ e^{-\im\varphi}\over \hbar}\int \diff t_{\varphi}{1\over 2}\left({\diff z\over \diff t_{\varphi}}\right)^2
\label{eq:wick02}
\ee
as a holomorphic functional of $z:[t_i,t_f]\to\mathbb{C}$ with $z(t_i)=x_i$ and $z(t_f)=x_f$. 
Since critical points are determined by the same Euler--Lagrange equation (\ref{eq:free02}), we can obtain them as $z_{\mathrm{cl}}(t_{\varphi})$ of (\ref{eq:free03}). 
By repeating the same procedure in (\ref{eq:free07}--\ref{eq:free11}), the Lefschetz thimble $\mathcal{J}$ around $z_{\mathrm{cl}}(t_{\varphi})$ can be denoted as 
\be
z(t_{\varphi})=z_{\mathrm{cl}}(t_{\varphi})+e^{\im\varphi/2}\sum_{\ell>0}a_{\ell}\sin\left(\pi \ell{t_{\varphi}-t_i\over t_f-t_i}\right)
\label{eq:wick03}
\ee
with $(a_{\ell})\in \ell^2(\mathbb{R})$. At $\varphi={\pi\over 2}$, this coincides with elements of Lefschetz thimbles for real-time path integral in (\ref{eq:free11}). Therefore, our result in (\ref{eq:free06}) and (\ref{eq:free15}) is naturally connected to previous studies of path integrals via imaginary-time by smoothly deforming Lefschetz thimbles as in (\ref{eq:wick03}) along the Wick rotation. 

\subsection{Free particle on a circle $S^1$}\label{sec:free_circle}
In this subsection, quantum mechanics of a free particle on a circle $S^1$ will be considered. The \schroedinger equation (\ref{eq:free_schr}) is the same, but we need to specify the boundary condition for $x\in S^1=\mathbb{R}/2\pi\mathbb{Z}$. We require that wave functions obey $\psi(x+2\pi,t)=e^{\im\theta}\psi(x,t)$ with a real parameter $\theta$, and we will compute the fundamental solution of this system. 

In this case, the exponent of the integrand in (\ref{eq:feynman}) is given by,
\be
\mathcal{I}[z]={\im} \int \diff t {1\over 2}\left({\diff z\over \diff t}\right)^2+\im{\hbar\theta\over 2\pi} \int \diff z,
\label{eq:circle01}
\ee
which must be regarded as a holomorphic functional of $z:\mathbb{R}\to \mathbb{C}/2\pi \mathbb{Z}$ with the boundary condition $z(t_i)=x_i$ and $z(t_f)=x_f$. 
The last term in (\ref{eq:circle01}) is called a topological term, which distinguishes the first homotopy class $\pi_1(S^1)$ of possible paths. 
The critical point is given by the same Euler--Lagrange equation (\ref{eq:free02}). 
Combining this with the boundary condition $z(t_i)=x_i$ and $z(t_f)=x_f$ in $\mathbb{R}/2\pi\mathbb{Z}\subset \mathbb{C}/2\pi\mathbb{Z}$, we find the set of solutions $\{z_{\mathrm{cl},w} (t)\}_{w\in\mathbb{Z}}$ which are given by 
\be
z_{\mathrm{cl},w}(t)=(x_f+2\pi w-x_i) {t-t_i\over t_f-t_i}+x_i. 
\label{eq:thimble03}
\ee
Let $\mathcal{J}_w$ be the Lefschetz thimble associates each classical solution $z_{\mathrm{cl},w}(t)$, which can be calculated as (\ref{eq:free11}). Path integral (\ref{eq:feynman}) of this system can be rewritten as 
\be
K_{\mathrm{free},\theta}(x_f,t_f;x_i,t_i)=\sum_{w\in\mathbb{Z}}\int_{\mathcal{J}_w}\Diff z \exp{\mathcal{I}[z]\over \hbar} =\sqrt{1\over 2\pi\im\hbar (t_f-t_i)}\sum_{w=-\infty}^{\infty}\exp{\mathcal{I}[z_{\mathrm{cl},w}]\over \hbar}. 
\label{eq:thimble06}
\ee
According to the formula (\ref{app:PicardLefschetz}), we need to sum up all classical solutions labeled by the topological number $w$. 

\subsection{Harmonic oscillator}\label{sec:ho}
So far, we have considered path integrals of free particles. Let us consider quantum mechanics of a harmonic oscillator in this formalism. The action functional of this system is given by 
\be
\mathcal{I}[z]={\im} \int \diff t \left[{1\over 2}\left({\diff z\over \diff t}\right)^2-{1\over 2}z^2\right],
\label{eq:ho01}
\ee
which is a holomorphic functional on $\mathcal{X}$. 
The Euler--Lagrange equation is 
\be
{\diff^2 z(t)\over \diff t^2}=-z(t). 
\label{eq:ho02}
\ee
Therefore, the classical solution of this differential equation is obtained as 
\be
z_{\mathrm{cl}}(t)={x_f-x_i\cos (t_f-t_i)\over \sin (t_f-t_i)}\sin(t-t_i)+x_i \cos (t-t_i). 
\label{eq:ho03}
\ee
In the following, we assume that the initial and final times, $t_i$ and $t_f$, are generic so that (\ref{eq:ho03}) makes sense for arbitrary $x_i,x_f\in\mathbb{R}$, that is, $t_f-t_i\not=n\pi$ for any $n\in\mathbb{Z}_{>0}$. The classical action of this solution is given as 
\be
\mathcal{I}[z_{\mathrm{cl}}]={\im \over 2\sin (t_f-t_i)}\left[(x_f^2+x_i^2)\cos (t_f-t_i)-2 x_f x_i\right]. 
\ee

The downward flow equation emanating from $z_{\mathrm{cl}}$ is given by 
\be
{\p \over \p u}\Delta z(t;u)=-\im \left({\p^2\over \p t^2}+1\right)\overline{\Delta z(t;u)}, 
\label{eq:ho04}
\ee
where $z(t;u)=z_{\mathrm{cl}}(t)+\Delta z(t;u)$ with boundary conditions $\Delta z(t;-\infty)=0$ and $\Delta z(t_i;u)=\Delta z(t_f, u)=0$. 
The set of solutions $\Delta z$ are spanned by 
\be
\Delta z_n(t;u)=\left\{\begin{array}{cc}
e^{\im\pi/4}\exp\left[\left(\left({\pi n\over t_f-t_i}\right)^2-1\right)u\right]\sin {n\pi\over t_f-t_i}(t-t_i),& (n\pi>(t_f-t_i)),\\
e^{-\im\pi/4}\exp\left[\left(1-\left({\pi n\over t_f-t_i}\right)^2\right)u\right]\sin {n\pi\over t_f-t_i}(t-t_i),& (n\pi<(t_f-t_i)). 
\end{array}\right.
\label{eq:ho05}
\ee
Let $\nu$ be the maximal non-negative integer smaller than $(t_f-t_i)/\pi$, then an element of the Lefschetz thimble $\mathcal{J}$ can be denoted as 
\be
z(t)=z_{\mathrm{cl}}(t)+e^{-\im\pi/4}\sum_{\ell=1}^{\nu}a_{\ell}\sin{\pi\ell\over t_f-t_i}(t-t_i)+e^{\im\pi/4}\sum_{\ell=\nu+1}^{\infty}a_{\ell}\sin {\pi \ell\over t_f-t_i}(t-t_i), 
\label{eq:ho06}
\ee
with $(a_{\ell})\in\ell^2(\mathbb{R})$. Therefore, functional integration measure on the Lefschetz thimble $\mathcal{J}$ becomes 
\be
\int_{\mathcal{J}}\Diff z=\mathcal{N}\int \prod_{n=1}^{\nu}e^{-{\im \pi\over 4}}\diff a_{n}\prod_{m=\nu+1}^{\infty}e^{\im \pi\over 4}\diff a_m=e^{-\im\pi\nu/2}\mathcal{N}\int \prod_{\ell=1}^{\infty}\sqrt{i}\diff a_{\ell}, 
\ee
which has an extra factor $\exp\left(-\im {\pi\nu\over 2}\right)$ compared with its counterpart (\ref{eq:free13}) for a free particle. The integer $\nu$ is Maslov--Morse index of the classical trajectory (see Ref.~\cite[Appendix 11]{arnold_mechanics}, for example), which represents the number of turning points in the time $t_f-t_i$. 

The Feynman kernel for the harmonic oscillator can now be readily calculated in the following way: 
\bea
K_{\mathrm{h.o.}}(x_f,t_f;x_i,t_i)&=&\exp\left({\mathcal{I}[z_{\mathrm{cl}}] \over \hbar}-\im{\pi\nu\over 2}\right) \mathcal{N}\prod_{\ell}\int \sqrt{i}\diff a_{\ell}\exp\left(-\left|\left({\pi \ell\over t_f-t_i}\right)^2-1\right|a_{\ell}^2\right) \nonumber\\
&=&\exp\left({\mathcal{I}[z_{\mathrm{cl}}] \over \hbar}-\im{\pi\nu\over 2}\right)\sqrt{1\over 2\pi\im\hbar(t_f-t_i) }\prod_{\ell=1}^{\infty}\sqrt{1\over |1-((t_f-t_i)/\pi\ell)^2|} \nonumber\\
&=&\sqrt{1\over 2\pi\im\hbar|\sin (t_f-t_i)| }\exp\left({\mathcal{I}[z_{\mathrm{cl}}] \over \hbar}-\im{\pi\nu\over 2}\right). 
\eea
In this calculation, we used the same normalization factor $\mathcal{N}$ calculated for the free particle. 
This calculation shows that Maslov index represents how the Lefschetz thimble intersects with the original space of paths, and its appearance becomes crystal--clear in this formulation. 

\section{Tunneling in real-time path integral}\label{sec:tunneling}
Tunneling is a universal phenomenon of quantum mechanics, and it has a fruitful application to realistic physics, such as decays of false vacua, bubble nucleation at first order phase transitions, domain wall fusions, and so on \cite{PhysRevD.15.2929, voloshin1975bubbles, Coleman198802, Shifman201201}. 
In order to study real-time dynamics of tunneling phenomena, we apply the Picard--Lefschetz theory to the path integral of quantum system with quasi-stable states.  

\subsection{Double well potential}
Let us pick up quantum mechanics in a double well potential. The classical action is given by 
\be
\mathcal{I}[z]=\im \int \diff t \left[{1\over 2}\left({\diff z\over \diff t}\right)^2-{1\over 2}(z^2-1)^2\right]. 
\label{eq:dw01}
\ee
The Euler--Lagrange equation of (\ref{eq:dw01}) is given by 
\be
{\diff^2 z\over \diff t^2}=-2z(z^2-1), 
\label{eq:dw02}
\ee
with the boundary condition $z(t_i)=x_i$ and $z(t_f)=x_f$. Instead of solving (\ref{eq:dw02}) directly, we consider the energy conservation, 
\be
\left({\diff z\over \diff t}\right)^2 + (z^2-1)^2=p^2, 
\label{eq:dw03}
\ee
with a complex constant $p\in\mathbb{C}$. The solution of this ordinary differential equation (\ref{eq:dw03}) can be explicitly written down using Jacobian elliptic functions: 
\be
z(t)=\sqrt{p^2-1\over 2p}\mathrm{sd}\left(\sqrt{2p}\;t+c,\sqrt{1+p\over 2p}\right), 
\label{eq:dw04}
\ee
where $c$ is an integration constant so that $z(t_i)=x_i$, and $k=\sqrt{(1+p)/2p}$ is called an elliptic modulus. Here we obey the notation given in Ref.~\cite[Chap.22]{NIST:DLMF}. 
\begin{figure}[t]
\centering
\includegraphics[scale=0.4]{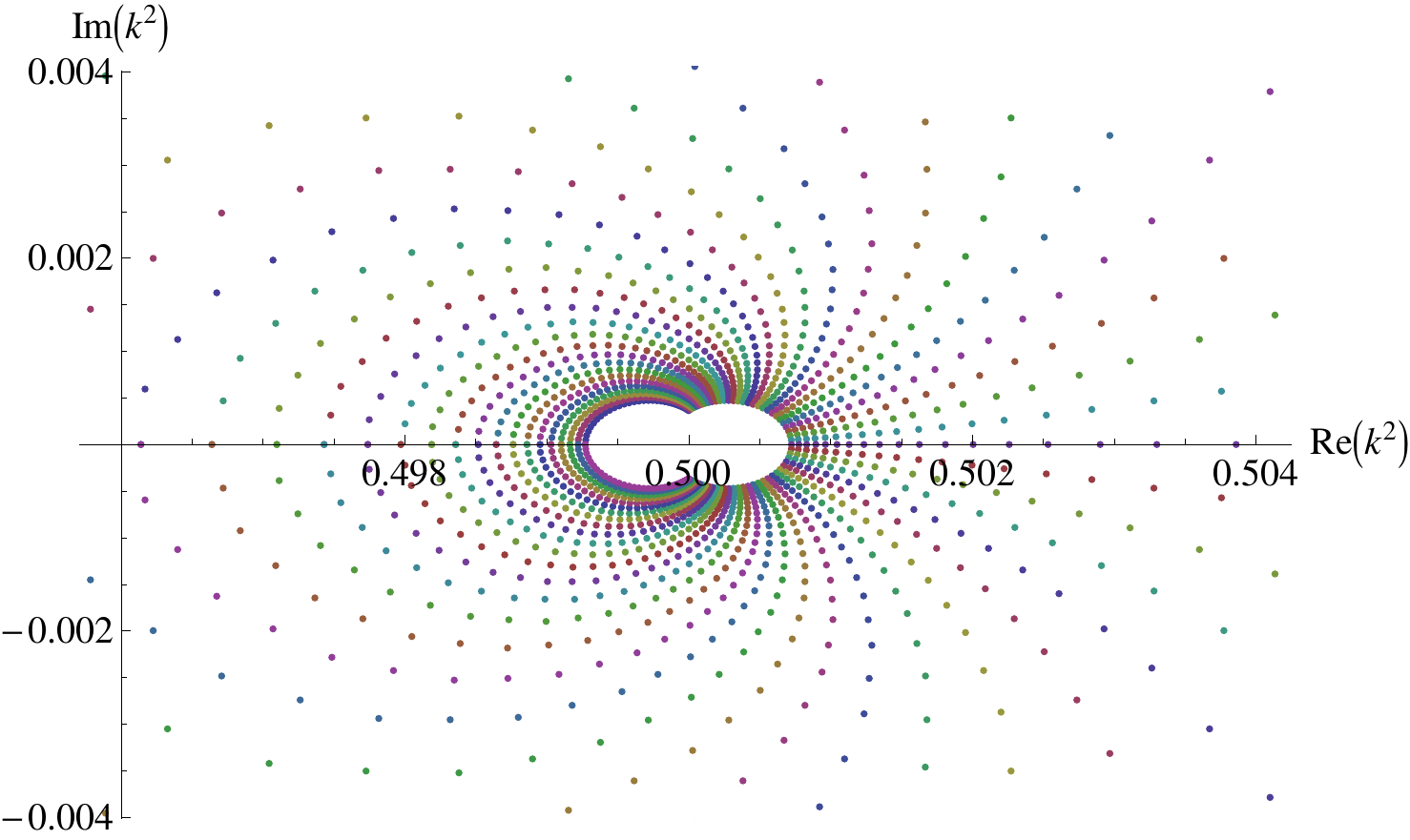}\qquad
\includegraphics[scale=0.4]{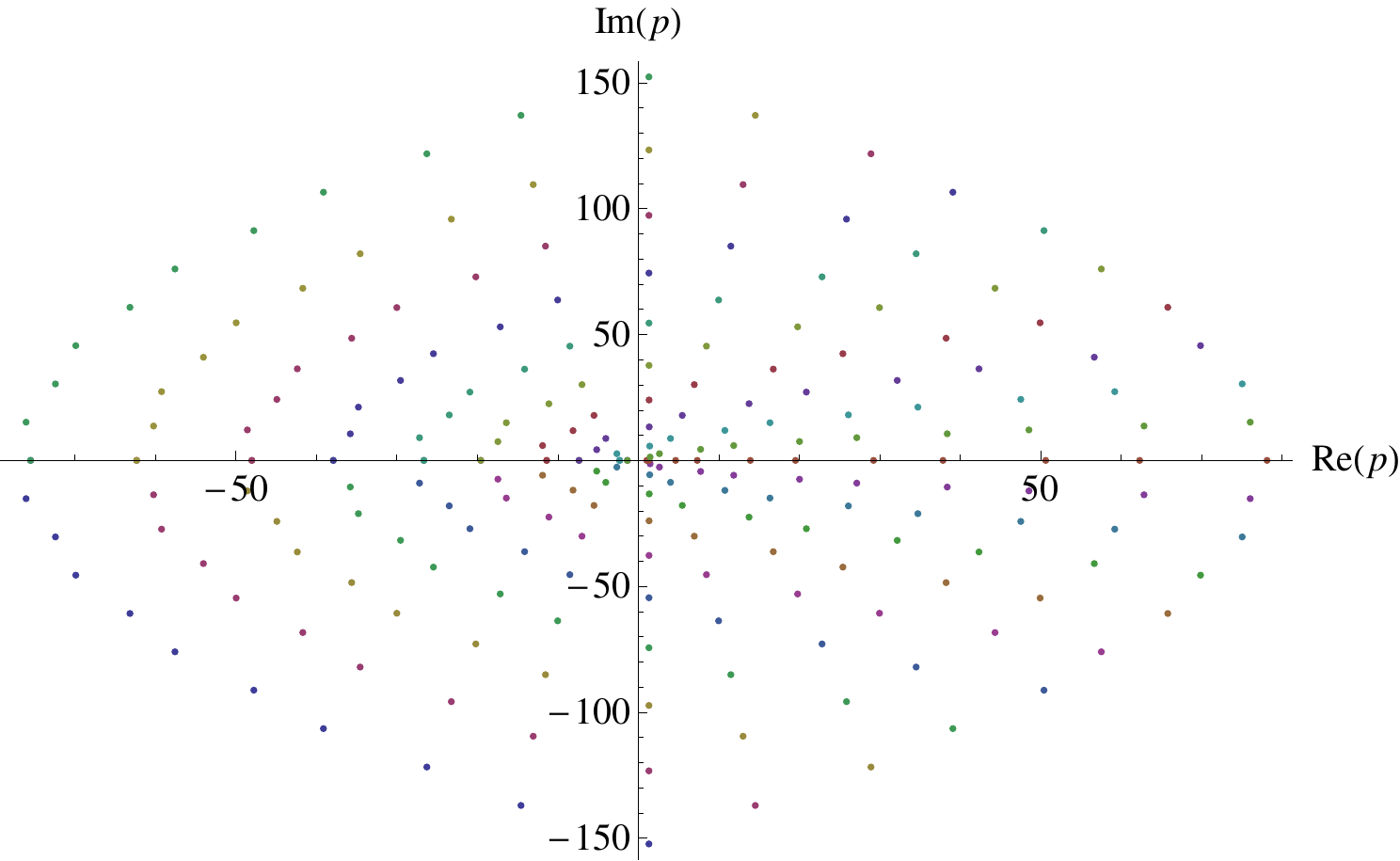}
\caption{List of classical solutions (\ref{eq:dw04}) with the boundary condition $x_i=-1$ and $x_f=1$: Left and right panels plot the list of parameters in complex $k^2$ and $p$ planes, respectively. These plots of parameters $k^2$ and $p$ are obtained by numerically solving the formula (\ref{eq:relation_classical_solutions_integers02}). }
\label{fig:list_k2}
\end{figure}
In order to calculate the set of the solutions of Euler--Lagrange equation (\ref{eq:dw02}), we must find all possible parameters $p$ satisfying $z(t_f)=x_f$. 
In Fig.~\ref{fig:list_k2}, a partial list of possible $k^2$ (and $p=1/(2k^2-1)$) is shown for the case $x_i=-1$ and $x_f=1$ and $t_f-t_i=3$, as an example. 
This problem will be considered in Section~\ref{sec:label_dw_solutions}, and we can see there why the figure shows a significant lattice structure. 

\begin{figure}[t]
\centering
\includegraphics[scale=0.65]{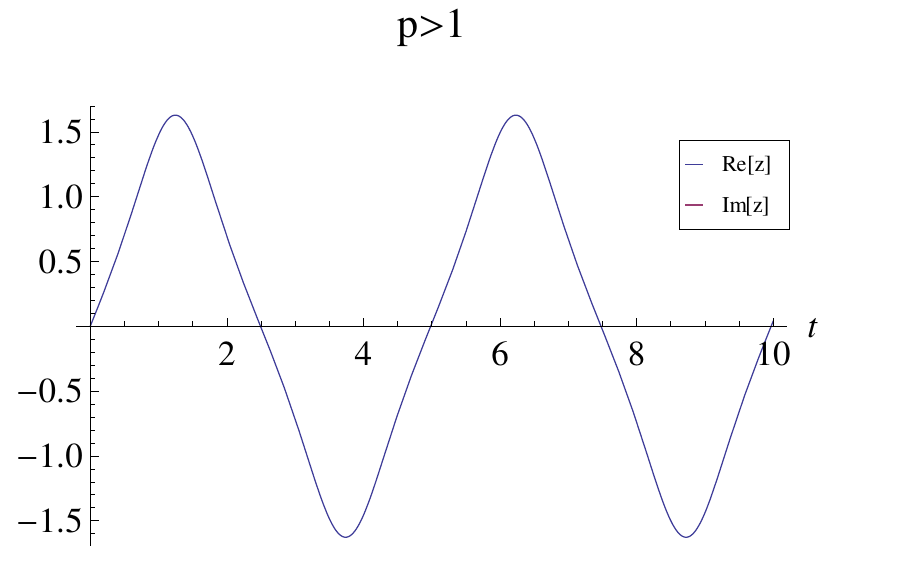}\quad
\includegraphics[scale=0.65]{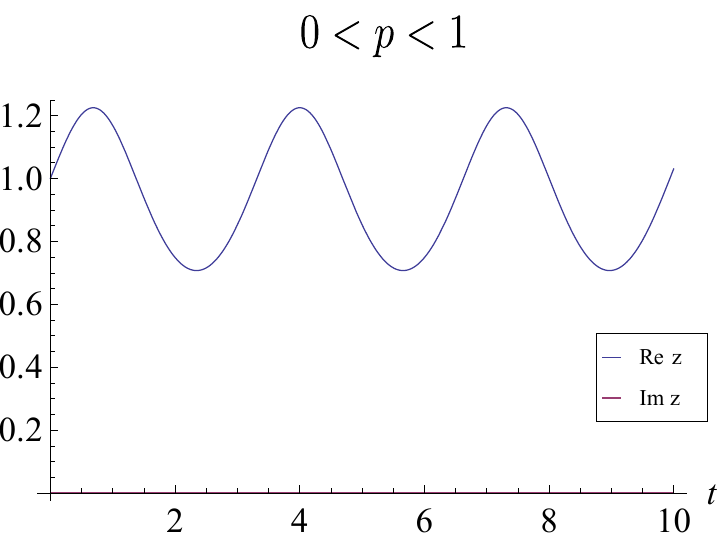}
\caption{Typical behaviors of real classical solutions. The left and right panels show those of $p>1$ and $p<1$, respectively. }
\label{fig:real_sol}
\end{figure}
Before computing the set of classical solutions, let us remark qualitative behaviors of each classical solution. 
Since the potential in (\ref{eq:dw01}) has two minima at $\pm 1$, there exist two different kinds of real classical solutions. If the total energy $p^2>1$, the particle can cross the barrier at the origin $0$ and it oscillates as shown in the left panel of Fig.~\ref{fig:real_sol}. If $0<p^2<1$, the particle is trapped in one of the minima as shown in the right panel of Fig.~\ref{fig:real_sol}. 
As we will see in the  next subsection, periods of these solutions can be characterized by complete elliptic integrals of the first kind. 

\begin{figure}[t]
\centering
\includegraphics[scale=0.6]{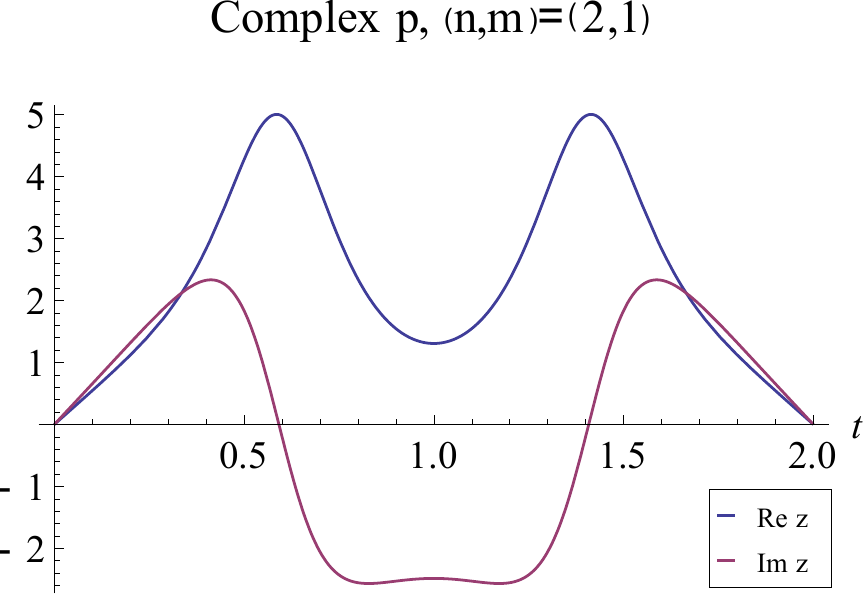}\quad
\includegraphics[scale=0.6]{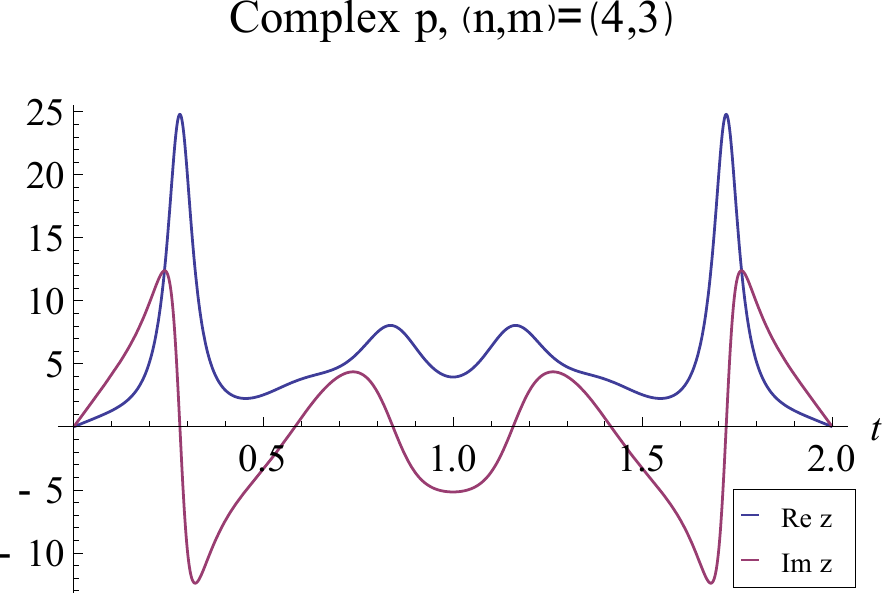}
\caption{Typical behaviors of complex classical solutions. $(n,m)$ refers an element of $\Sigma$ constructed in the next subsection. }
\label{fig:cplx_sol}
\end{figure}
So far, we have discussed properties of real classical solutions, but there also exists complex solutions. 
Fig.\ref{fig:cplx_sol} shows typical behaviors of complex solutions of (\ref{eq:dw02}). 
Thanks to these complex solutions, there can exist tunneling solution connecting between two different potential minima without exceeding the potential barrier. This must be compared with the fact that there does not exist such real classical solutions with energies lower than the potential barrier. 
This point will be discussed in more detail in the context of quantum tunneling and symmetry restoration in Section~\ref{sec:discussion_tunneling}. 

\subsection{Classical solutions}\label{sec:label_dw_solutions}

\subsubsection{Classification of classical solutions}
Fix real numbers $t_i <t_f$ and denote by $\Sigma_{x_i, x_f}$ the set
\begin{equation}\left\{z\colon [t_i, t_f]\to \mathbb{C}\left| \ \frac{\mathrm{d}^2z}{\mathrm{d}t^2}=-2z(z^2-1),\ z(t_i)=x_i,\ z(t_f)=x_f\right. \right\}, \end{equation}
of classical solutions for an initial point $x_i\in\mathbb{C}$ and a final point $x_f\in\mathbb{C}$. 
The goal of this subsection is to classify classical solutions when $x_i=x_f=0$. 
More precisely, we will construct the one-to-one correspondence between $\Sigma_{0, 0}/\{\pm1\}$ and the set 
\begin{equation}
\Sigma:=\left\{[(n, m)]\in\mathbb{Z}^2/\sim\left| \,\frac{n}{{\rm gcd}(n, m)}\cdot\frac{m}{{\rm gcd}(n, m)}\equiv 0\ \text{mod}\ 2\right.\right\}
\end{equation}
where $\text{gcd}(n, m)$ is the greatest common divisor of $n$ and $m$ and 
$\sim$ is a equivalence relation generated by $(n, m)\sim (-n, -m)$. 
Here we formally regard $\text{gcd}(n, m)$ as the number $1$ when $nm=0$ holds. 

\subsubsection{Choice of branches of elliptic integrals}
In Section~\ref{sec:label_dw_solutions},  we use functions $k\mapsto\sqrt{k^2-1/2}K(k)$ and $k\mapsto\mathrm{i}\sqrt{k^2-1/2}K(\sqrt{1 - k^2})$, where $K(\ell)$ is complete elliptic integral of the first kind $\int_0^{\frac{\pi}{2}}(1-\ell^2\sin^2x)^{-\frac{1}{2}}\,\mathrm{d}x$. 
Let us fix a branch of the function $k\mapsto\sqrt{(2k^2-1)/2}K(k)$ and denote this single-valued function by $\omega_1(k)$. 
As $\omega_3(k):=\omega_1(\sqrt{1-k^2})=\pm\mathrm{i}\sqrt{(2k^2-1)/2}K(\sqrt{1 - k^2})$, we can choose the branch of the function $k\mapsto\mathrm{i}\sqrt{(2k^2-1)/2}K(\sqrt{1 - k^2})$ such that 
\begin{equation}\label{choiceofbranches}
\omega_3(k)=\mathrm{i}\sqrt{{2k^2-1\over 2}}K(\sqrt{1 - k^2})
\end{equation}
holds. In this section~\ref{sec:label_dw_solutions}, we always use such a choice of branches. 
The branch cut of $n\omega_1+m\omega_3$ consists of two half lines $(-\infty,1/2]\cup [1,\infty)$ in the complex $k^2$ plane. 

In this section, we first show that there is one-to-one correspondence between $\Sigma_{0,0}/\{\pm1\}$ and $\Sigma$ by 
\begin{equation}
n\omega_1(k)+m\omega_3(k)=\frac{t_f-t_i}{2}, 
\label{eq:relation_classical_solutions_integers01}
\end{equation}
with (\ref{eq:dw04}). We also consider its generalization to an arbitrary boundary condition in Section~\ref{subsec:general_boundary_conditions}. 

\subsubsection{Construction of the correspondence $\Sigma_{0, 0}/\{\pm1\}\to \Sigma$}
Here we construct the map $\Sigma_{0, 0}/\{\pm1\}\to \Sigma$. 
First we define that the image of the element $0$, the classical solution which is identically equal to $0$, is $(0, 0)$. 

Next, we will define the image of an element $z\in\Sigma_{0, 0}\setminus\{0\}$. 
As we mentioned, $z$ satisfies (\ref{eq:dw03}) for some constant $p\in\mathbb{C}$. 
In the case where $p^2=1$, we can conclude by simple computation that $z\equiv 0$ holds, which corresponds to $[(0,0)]\in\Sigma$ by the above definition. 
When $p=0$, we can deduce from (\ref{eq:dw03}) that $z=\im \tan (t+c)$ with some $c$ or that $z(t)=\pm 1$. 
As those functions cannot be a classical solution with the given boundary condition, we can conclude that $p\in\mathbb{C}\setminus\{0, \pm1\}$ in the following. 

Let $X\colon [t_i, t_f]\to\mathbb{C}$ be a function defined by $X:=-z^2+2/3$. 
Then $X$ enjoys the differential equation 
\begin{equation}\left(\frac{\mathrm{d}X}{\mathrm{d}t}\right)^2=4\left(X-\frac{2}{3}\right)\left(X-\left(p-\frac{1}{3}\right)\right)\left(X+\left(p+\frac{1}{3}\right)\right), 
\end{equation}
and thus $X$ can be extended to a meromorphic function $X$ defined on $\mathbb{C}$ which can be written in the form  
\begin{equation}X(t)=\wp(t+C, \Lambda_k)\end{equation}
by using a Weierstrass elliptic function $\wp(-, \Lambda_k)$ with the period lattice $\Lambda_k$ generated by $\omega_1(k)$ and $\omega_3(k)$, where $C\in\mathbb{C}$ is a constant and $k=\sqrt{({p+1})/{2p}}$ (see Ref.~\cite[Eq.(23.6.16)]{NIST:DLMF}). 
Let us denote  $\omega_1+\omega_3$ by $\omega_2$. 
\begin{figure}[t]
\centering
\includegraphics[scale=0.2]{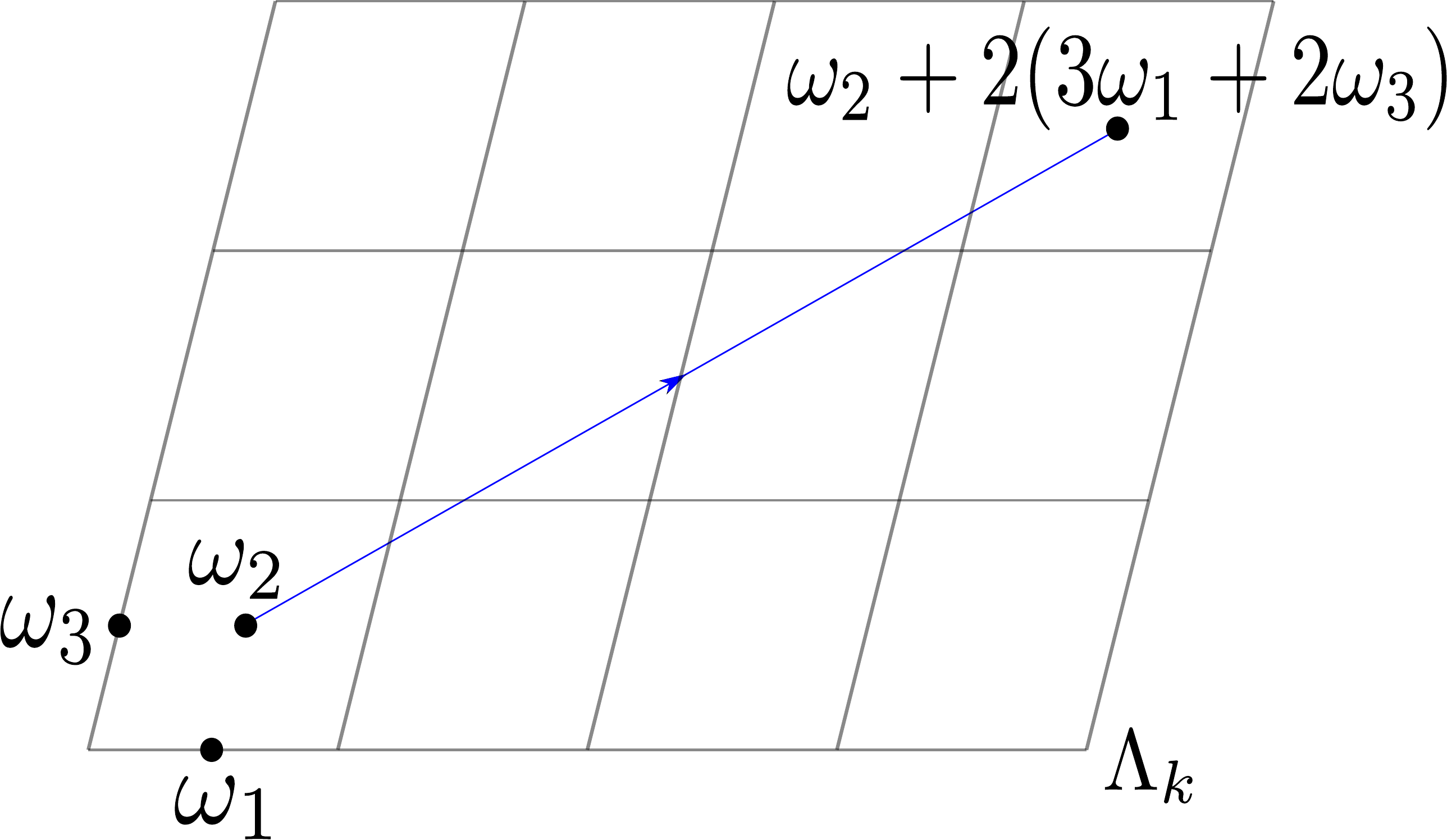}
\caption{$[t_i+C, t_f+C]$ as a line segment embedded in $\mathbb{C}$. The classical solution labeled by $(n,m)\in \Sigma$ goes straight from $\omega_2$ to $\omega_2+2n\omega_1+2m\omega_3$ during the time $t_f-t_i$ without intersecting the lattice $\Lambda_k$. }
\label{fig:real_lattice}
\end{figure}
According to Ref.~\cite[Sec. 23.3(i)]{NIST:DLMF}, $\wp(t+C, \Lambda_k)$ satisfies $\wp(\omega_1, \Lambda_k)=p-\frac13$,  $\wp(\omega_2, \Lambda_k)=\frac23$, and $\wp(\omega_3, k)=-p-\frac13$. We here remark that $\omega_2$ is the unique zero of the function $\wp(-, \Lambda_k)-\frac23$ up to the period $\Lambda_k$  and that $\wp$ diverges only at the origin $0$ up to $\Lambda_k$. 
Thus it is clear that both $t_i+C$ and $t_f+C$ are elements of the lattice $\omega_2+\Lambda_k$ (see Fig.\ref{fig:real_lattice}). 
Therefore, we can take a pair of integers $(n, m)$ such that (\ref{eq:relation_classical_solutions_integers01}) holds. 
We define the map $\Sigma_{0, 0}/\{\pm1\}\to \mathbb{Z}^2/\sim$ by $z\mapsto [(n, m)]\in\mathbb{Z}^2/\sim$ 
(It is well-defined since we choose branches such that the equation (\ref{choiceofbranches}) holds).

Finally we have to check that this $(n, m)$ is an element of $\Sigma$, or equivalently, that $\frac{n}{{\rm gcd}(n, m)}\cdot\frac{m}{{\rm gcd}(n, m)}$ is an even integer. 
It is clear because, as the function $X=-z^2+2/3$ has no pole on the line segment $[t_i+C, t_f+C]$, we can deduce that the line segment $[t_i+C, t_f+C]$ does not intersects the lattice $\Lambda_k$ (Fig.\ref{fig:real_lattice} shows this fact for the case $(n,m)=(3,2)$). 

\subsubsection{Construction of the inverse map $\Sigma\to\Sigma_{0, 0}/\{\pm1\}$}

From now on, we will construct the inverse correspondence. 
Fix an element $[(n, m)]\in\Sigma$. 

First we will show the existence of a constant $k\in\mathbb{C}\setminus\{0, \pm(1/\sqrt{2}), \pm1\}$ 
which satisfies (\ref{eq:relation_classical_solutions_integers01}).  
Let us consider a set $C:=\{k^2\in\mathbb{C}\mid\mathrm{Im}(n\omega_1+m\omega_3)(k)=0\}$. 
It is clear that, as a subset of $\mathbb{CP}^1={\mathbb{C}}\cup\{\infty\}$, $C$ is a real $1$-dimensional smooth connected curve with two boundary points. Since $\omega_1=\omega_3=0$ at $k^2=1/2$, one of the endpoints is $1/2$.

\begin{figure}[t]
\centering
\includegraphics[scale=0.28]{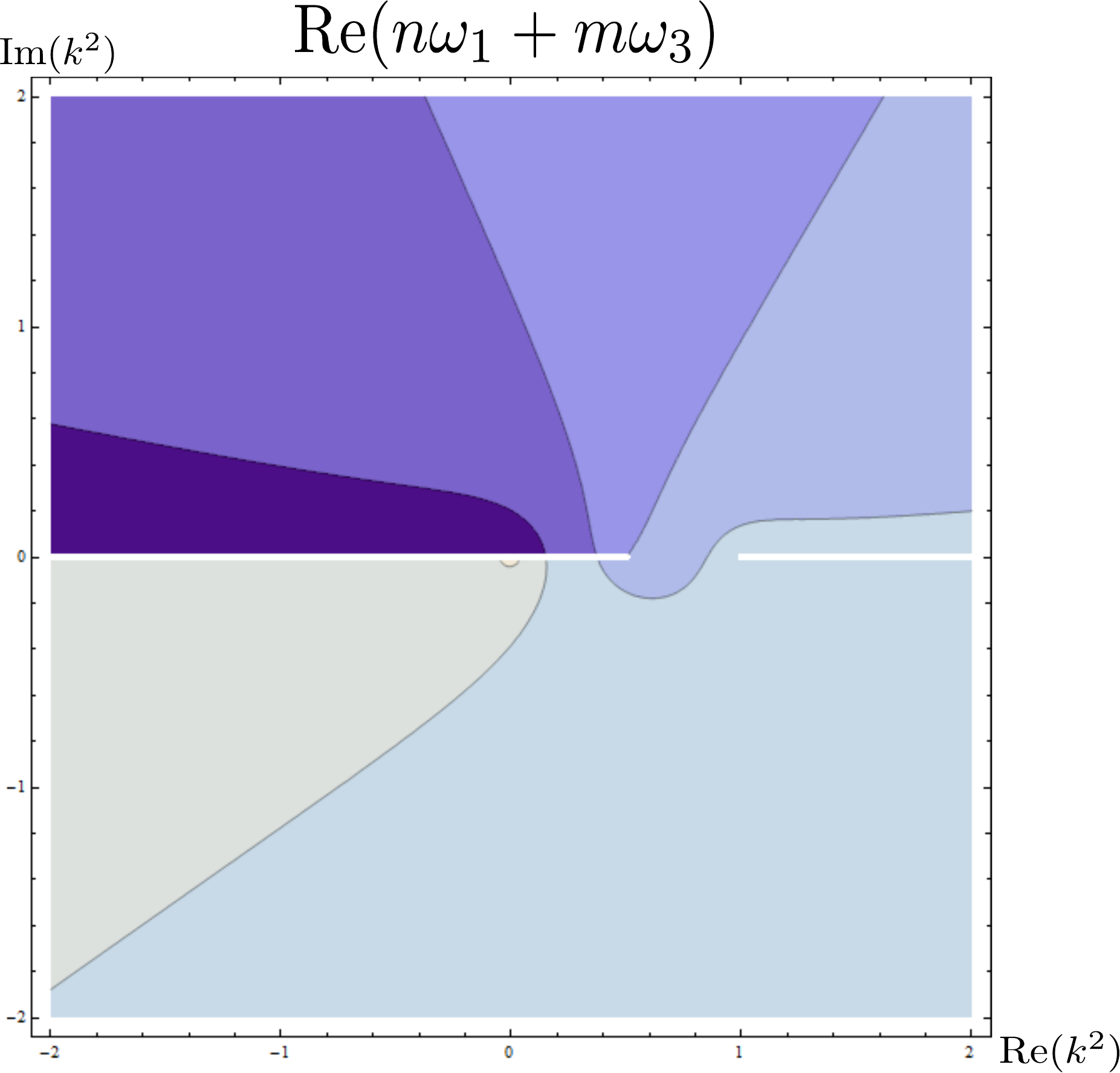}\qquad
\includegraphics[scale=0.28]{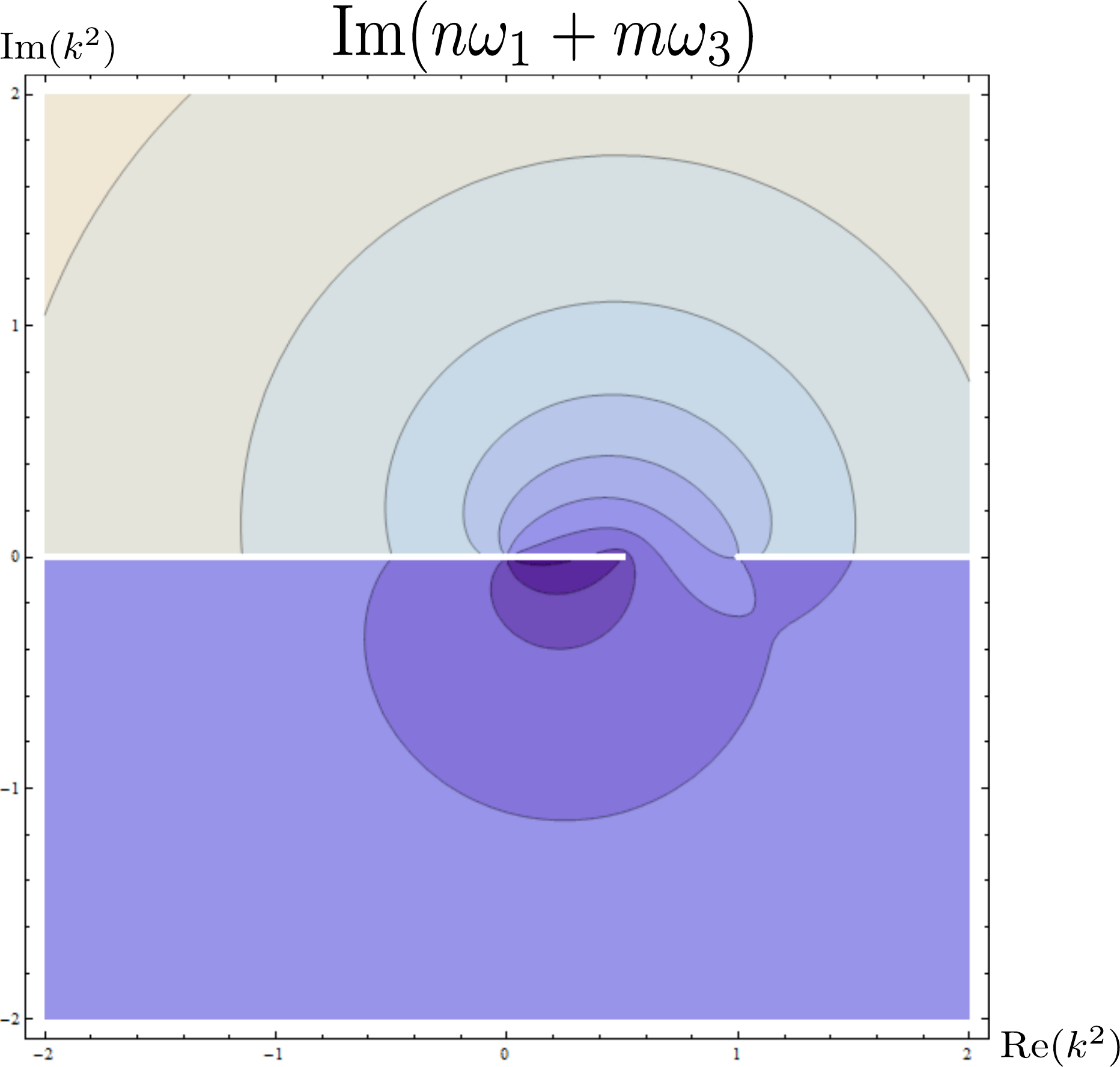}
\caption{Contour plots for real and imaginary parts of $n\omega_1+m\omega_3$ as functions of $k^2$. The branch cut $(-\infty,1/2]\cup[1,\infty)$ is cut off from these plots. }
\label{fig:cont_plot01}
\end{figure}

Let us show that another boundary point of $C$ is $0$ or $1$. We introduce a function $f=\mathrm{Re}(n\omega_1+m\omega_3)|_{C}$ on the curve $C$, which is a monotone function because each leaf of the foliation $\{k\in\mathbb{C}\mid\mathrm{Re}(n\omega_1+m\omega_3)(k)=\lambda\}_{\lambda\in\mathbb{R}}$ intersects $C$ transversally (see Fig.\ref{fig:cont_plot01} as an example). 
Therefore, another endpoint of $C$ must belongs to the branch cut $\mathbb{R}\setminus(1/2,1)$ of $n\omega_1+m\omega_3$. 
If $k^2>1$, we can calculate its imaginary part by using connection formulae given in Ref.~\cite[Sec.~19.~7]{NIST:DLMF}: 
\bea
\mathrm{Im}\left(n\omega_1(k)+m\omega_3(k)\right)&=&{\sqrt{2k^2-1\over 2k^2}}\mathrm{Im}\left(n K(1/k)+{\im}{(m \mp n)}K(\sqrt{1-1/k^2})\right)\nonumber\\
&=&\sqrt{2k^2-1\over 2k^2}(m\mp n)K(\sqrt{1-1/k^2}), 
\eea
which cannot be zero since $n\not=\pm m$ for $[(n,m)]\in\Sigma$. This proves that another endpoint of $C$ does not belong to $(1,\infty)$, and a similar computation shows that another endpoint must be $0$ or $1$. 
Since $K(k^2)\sim \ln 1/\sqrt{1-k^2}+\mathcal{O}(1)$ in the limit $k^2\to 1$, $f=\mathrm{Re}(n\omega_1+m\omega_3)$ diverges as $k^2\to 0$ or $1$. 
Therefore, it follows that there uniquely exists a point $\widetilde{k}\in C$ such that 
\begin{equation}
f(\widetilde{k})=\frac{t_f-t_i}{2}\ {\rm or}\ -\frac{t_f-t_i}{2}
\end{equation}
holds. We can set $k=\pm\sqrt{\widetilde{k}}$. 

Using this $k$, we can construct an element $z\in\Sigma_{0, 0}\setminus\{0\}$ by 
\begin{equation}
z(t)=\pm\sqrt{-\wp(t-t_i+\omega_2(k), \Lambda_k)+\frac23}, 
\label{eq:sol_weierstrass}
\end{equation}
where $\omega_2(k):=\omega_1(k)+\omega_3(k)$ and $\Lambda_k$ is a lattice generated by $2\omega_1(k)$ and $2\omega_3(k)$. 
We here remark that, as $k$ is unique up to a sign, the lattice $\Lambda_k$ is uniquely determined by the element $[(n, m)]\in\Sigma$. 
Since $[(n, m)]=[(-n, -m)]$, without loss of generality, we may assume that (\ref{eq:relation_classical_solutions_integers01}) holds. Then clearly the image of this $z$ is $[(n, m)]$ and it gives the inverse correspondence. 
The relation between two expressions (\ref{eq:dw04}) and (\ref{eq:sol_weierstrass}) of the solution can be found in Ref.~\cite[Eq.(23.6.25)]{NIST:DLMF}.

\subsubsection{Generalization to arbitrary boundary conditions}\label{subsec:general_boundary_conditions}

Let us consider the set of elliptic modulus $k$ for classical solutions with the boundary condition $z(t_i)=x_i$ and $z(t_f)=x_f$. Since the Weierstrass elliptic function describes $z_{\sigma}(t)^2$ instead of classical solutions themselves, it is cumbersome to describe the solutions with general boundary conditions. Therefore, we use the solution (\ref{eq:dw04}) represented by the Jacobian elliptic function. 
The Jacobian elliptic function obeys the following half-periodic properties \cite[Sec.~22.~4]{NIST:DLMF}: 
\be 
\mathrm{sd}(z+2K(k),k)=\mathrm{sd}(z+2iK(\sqrt{1-k^2}),k)=-\mathrm{sd}(z,k). 
\ee
We can now interpret $(n,m)$ in the relation (\ref{eq:relation_classical_solutions_integers01}) solves ambiguities of the inverse elliptic function due to the above half-periodicity. 
By solving the boundary condition ($k'=\sqrt{1-k^2}$)
\be
x_f={k k'\over \sqrt{(2k^2-1)/2}}\mathrm{sd}\left({t_f-t_i\over \sqrt{(2k^2-1)/2}}+\mathrm{sd}^{-1}\left({\sqrt{(2k^2-1)/2}\over kk'}x_i,k\right),k\right)
\ee
in terms of $k$, we could generalize (\ref{eq:relation_classical_solutions_integers01}) for general boundary conditions as 
\bea
&&n\omega_1(k)+m\omega_3(k)={t_f-t_i\over 2}\nonumber\\
&&+{\sqrt{2k^2-1}\over 2\sqrt{2}}\left(\mathrm{sd}^{-1}\left({\sqrt{2k^2-1}\over \sqrt{2}kk'} x_i,k\right)-(-1)^{n+m}\mathrm{sd}^{-1}\left({\sqrt{2k^2-1}\over \sqrt{2}kk'}x_f,k\right)\right). 
\label{eq:relation_classical_solutions_integers02}
\eea
We do not give a proof, however we conjecture that the similar argument will show the one-to-one correspondence between the set of classical solutions $\Sigma_{x_i,x_f}$ and an appropriate subset of $\mathbb{Z}^2/\sim$ by the relation (\ref{eq:relation_classical_solutions_integers02}). 

\subsection{Discussion}\label{sec:discussion_tunneling}
We discuss problems of Lefschetz-thimble technique on path integrals. In this part, we do not give a mathematically solid argument like in the previous subsection, but pose some important problems to be solved together with their physical motivations. 
\subsubsection{Intersection numbers $n_{\sigma}$ and short-time asymptotic behaviors}
We obtained the set of complex classical solutions in the previous section, and then we can compute the classical action at those saddle points. 
In the short-time limit $t_f-t_i\to +0$, we can easily calculate asymptotic behaviors of classical solutions, and those computations turn out to provide fruitful information on intersection numbers $n_{\sigma}$. 

For simplicity, let us again set $x_i=x_f=0$, and assume that $[(n,m)]\in \Sigma$. In order to consider short-time behavior of real-time solutions, we must solve (\ref{eq:relation_classical_solutions_integers01}) (or, more generally, (\ref{eq:relation_classical_solutions_integers02}))  in order to satisfy the boundary condition: 
\be
\sqrt{2k^2-1\over 2}\left(n K(k)+\im m K(\sqrt{1-k^2})\right)={t_f-t_i\over 2}. 
\label{eq:asymp01}
\ee
Since the right-hand-side goes to zero in the limit $t_f-t_i\to +0$, the elliptic modulus $k$ must converge to $1/\sqrt{2}$. This observation helps us to find asymptotic behavior of $k^2$ for given $(n,m)$: 
\be
k^2={1\over 2}+\left({t_f-t_i\over 2(n+\im m) K(1/\sqrt{2}) }\right)^2+\mathcal{O}\left((t_f-t_i)^4\right). 
\label{eq:asymp02}
\ee
As a result, one can find that the classical solution can be approximated as 
\be
z_{(n,m)}(t)\simeq {(n+\im m)K(1/\sqrt{2})\over t_f-t_i}\mathrm{sd}\left(2(n+\im m)K(1/\sqrt{2}){t-t_i\over t_f-t_i},{1\over \sqrt{2}}\right). 
\label{eq:asymp03}
\ee
This formula for the leading behavior in the limit $t_f-t_i \to 0$ implies that the solution is real if and only if $n=0$ or $m=0$, and other solutions become complex. Since the total energy of this solution $p^2/2$ behaves as 
\be
{p^2_{(n,m)}\over 2}={1\over 2(2k^2-1)^2}\simeq 2\left({(n+\im m)K(1/\sqrt{2})\over t_f-t_i}\right)^4, 
\label{eq:asymp04}
\ee
we can evaluate the asymptotic behavior of the classical action $\mathcal{I}$ only by taking into account the terms of the order of $\mathcal{O}(1/(t_f-t_i)^4)$ in the Lagrangian. Here, $p_{(n,m)}$ refers the parameter $p$ of the label $(n,m)$. Therefore, 
\bea
\mathcal{I}[z_{(n,m)}] &=& \im \int_{t_i}^{t_f}\diff t \left({p^2_{(n,m)}\over 2}-(z_{(n,m)}^2-1)^2\right) 
\simeq \im \int_{t_i}^{t_f}\diff t \left({p_{(n,m)}^2\over 2}-\wp(t+C,\Lambda_{k_{(n,m)}})^2\right)\nonumber\\
&=& \im{2K(1/\sqrt{2})^4\over 3}{(n+\im m)^4\over (t_f-t_i)^3}. 
\label{eq:asymp05}
\eea
In order to derive the last expression, we used a formula in Ref.~\cite[Sec.~23.14]{NIST:DLMF}. 
All these analyses work also for the imaginary-time formalism, and we just need to replace $t$, $t_i$, and $t_f$ by $-\im t$, $-\im t_i$, and $-\im t_f$, respectively. 

For real-time cases, the classical action $\mathcal{I}$ is purely imaginary on the original integration cycle, and thus complex saddles with $\mathrm{Re}\;\mathcal{I}>0$ do not contribute to the path integral. 
As already discussed, asymptotic behaviors (\ref{eq:asymp03}) of classical solutions with $n=0$ or $m=0$ are real functions, and then their intersection numbers $n_{\sigma}$ must be equal to one. 
We can conclude that there are already infinitely many saddle points contributing to the path integral for the double-well potential. 
If $n\not=0$ and $m\not=0$, all the solutions are necessarily complex and not real. In the following, we set $n>0$ to fix a representative of $[(n,m)]\in\Sigma$. According to (\ref{eq:asymp05}), the real part of $\mathcal{I}$ is approximately given by 
\be
\mathrm{Re}\;\mathcal{I}[z_{(n,m)}]\simeq -{8K(1/\sqrt{2})^4\over 3(t_f-t_i)^3}n m (n^2-m^2). 
\label{eq:asymp06}
\ee
Complex solutions with $nm(n^2-m^2)<0$ have positive real parts of the classical action, and their intersection numbers $n_{\sigma}$ must be zero. Therefore, complex solutions with $n<m$ or $n>-m>0$ do not contribute to the path integral. 
On the other hand, complex solutions with $n>m>0$ or $n<-m$ can, because they have negative real parts of the classical action. There still remain infinite complex solutions, whose integral coefficients $n_{\sigma}$ cannot be automatically determined. For those solutions $z_{\sigma}$, we need to solve upward flow equations and calculate intersection numbers $n_{\sigma}$ between $\mathcal{Y}$ and $\mathcal{K}_\sigma$. 

Calculation of undetermined coefficients $n_{\sigma}$ is an open problem. In order to obtain them, we must count the number of upward flows connecting $z_{\sigma}$ and some real paths in an appropriate way, and thus we need to know correct behaviors of nonlinear partial differential equations. 
This problem happens in general, and is very difficult to be solved. 
In order to see this fact, let us take lattice regularization of the path integral.  In a generic case with a polynomial potential, the  number of complex classical solutions are given by 
\be
\left(\mathrm{deg}V-1\right)^N, 
\ee
where $\mathrm{deg}V$ refers the degree of the potential term and $N$ does the number of lattice sites. 
Except for quadratic Lagrangians ($\mathrm{deg}V=2$), the classical equation of motion cannot be solved uniquely, and it has infinitely many solutions in the continuum limit $N\to \infty$. 
It then seems quite general that there are complex and not real saddle points, and half of them must have negative real parts of the classical action according the general constraint of real-time formalism discussed in Section~\ref{sec:basic}. 
Short-time asymptotic analysis is useful to evaluate the classical action as we have done in (\ref{eq:asymp01}-\ref{eq:asymp06}), and then we can show $n_{\sigma}=0$ for about half of classical solutions with $\mathrm{Re}\; \mathcal{I}[z_{\sigma}]>0$. 

We have no idea how to calculate $n_{\sigma}$ for the rest of complex solutions and leave this an open problem. We would like to emphasize that it is very important not only for theoretical interest on this formulation but also for practical applications of this formalism to sign problems. 
On the other hand, since this may provide a ``natural'' definition on intersections between infinite-dimensional manifolds motivated from physics computations, it seems to be interesting also from mathematical viewpoint. 

In the next part, we discuss importance of this problem for semi-classical treatment of real-time quantum tunneling. 

\subsubsection{Quantum tunneling}
Let us discuss quantum tunneling in the context of our computations of path integrals on Lefschetz thimble. 
Before doing it, let us recall physical consequence of quantum tunneling on this system. 
First of all, the ground state wave function $\psi_0$ becomes a parity even state. 
On the other hand, the first excited state $\psi_1$ is a parity odd state, and, in the semi-classical limit $\hbar\to 0$, $\psi_{\pm}=(\psi_0\pm\psi_1)/\sqrt{2}$ behaves as 
\be
\lim_{\hbar\to 0}|\psi_{\pm}(x)|^2=\delta(x\mp 1). 
\ee
That is, the localized state in one of the classical vacua $x=\pm 1$ is realized as a coherent summation of $\psi_0$ and $\psi_1$, and the oscillation between $\psi_{\pm}$ is nothing but the consequence of quantum tunneling. 
Therefore, $\psi_{\pm}$ fails to be eigenstates of the Hamiltonian, but its failure of stationarity cannot be described within the perturbation theory. Indeed, the energy difference is $E_1-E_0\sim \exp-S_0/\hbar$ with $S_0$ a positive constant. 
\begin{figure}[t]
\centering
\includegraphics[scale=0.35]{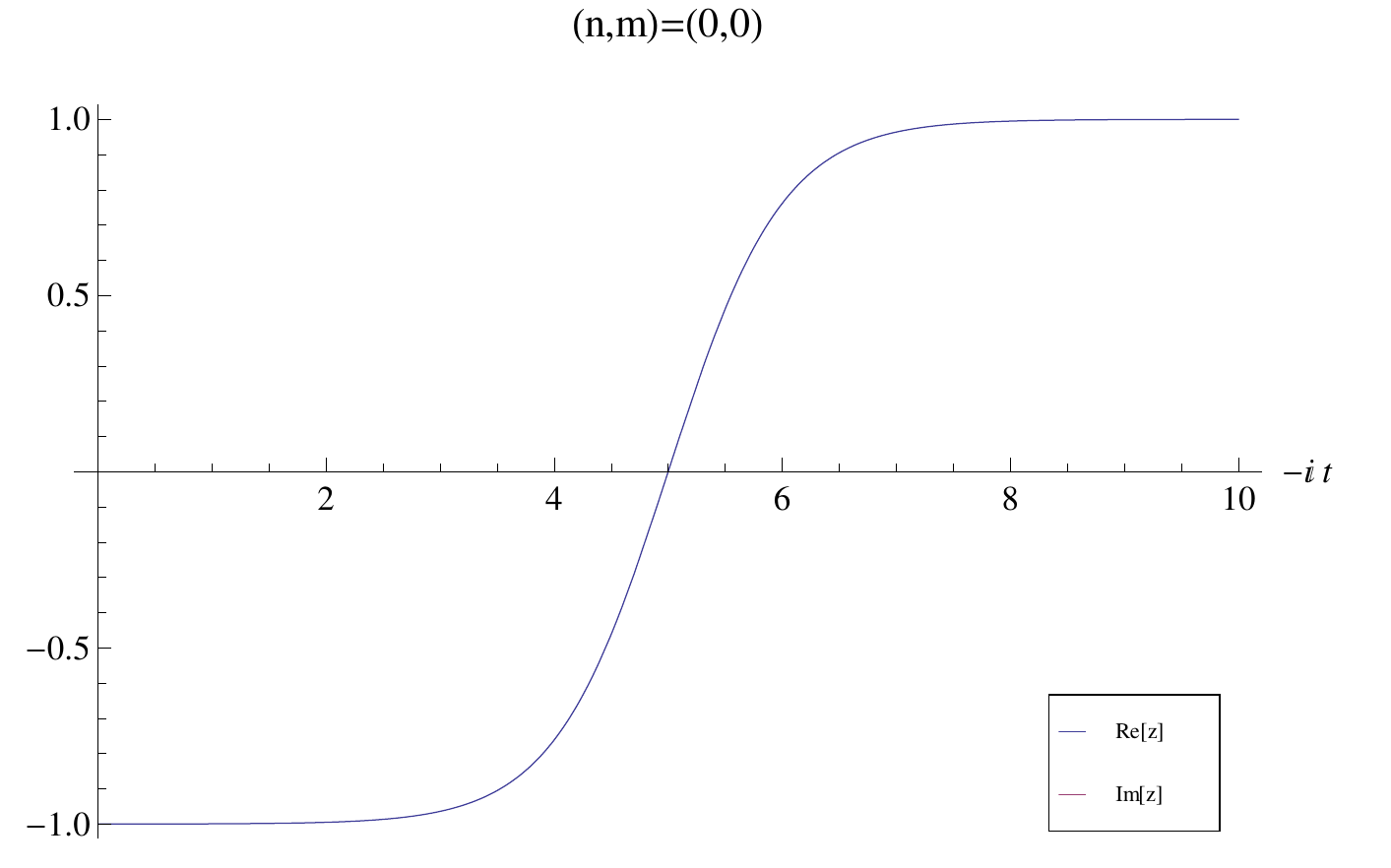}\;
\includegraphics[scale=0.35]{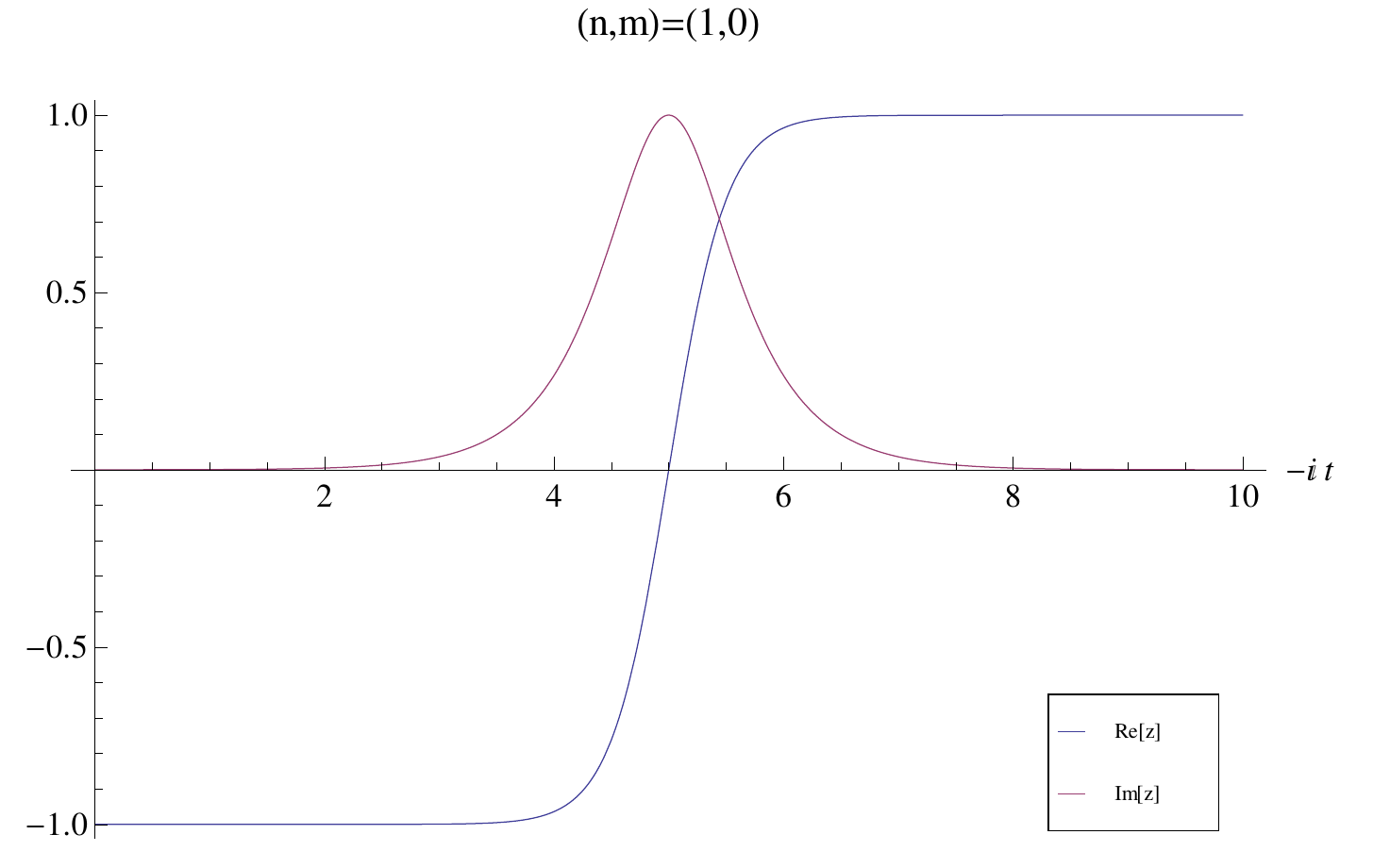}\;
\includegraphics[scale=0.35]{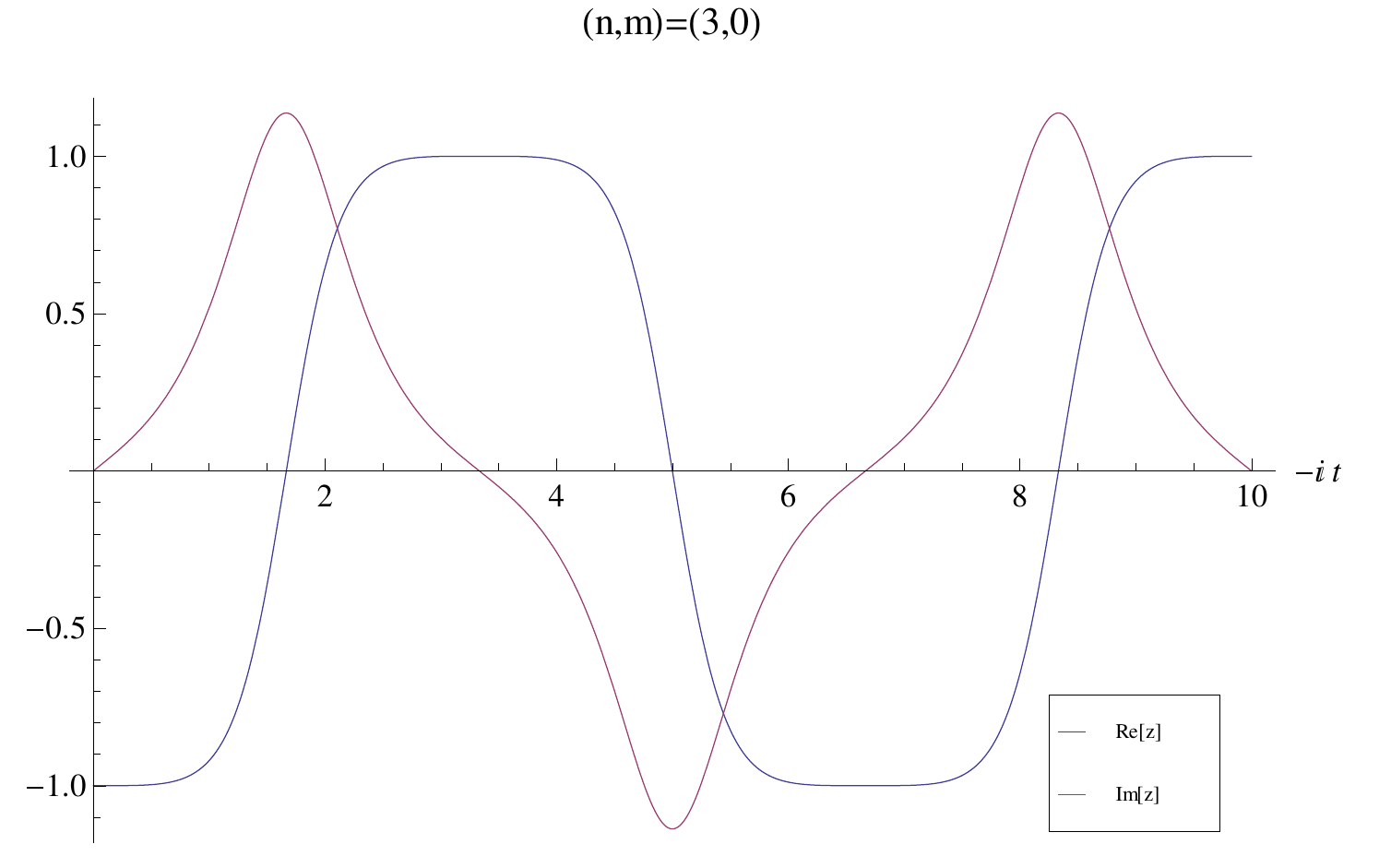}(a)\\
\includegraphics[scale=0.38]{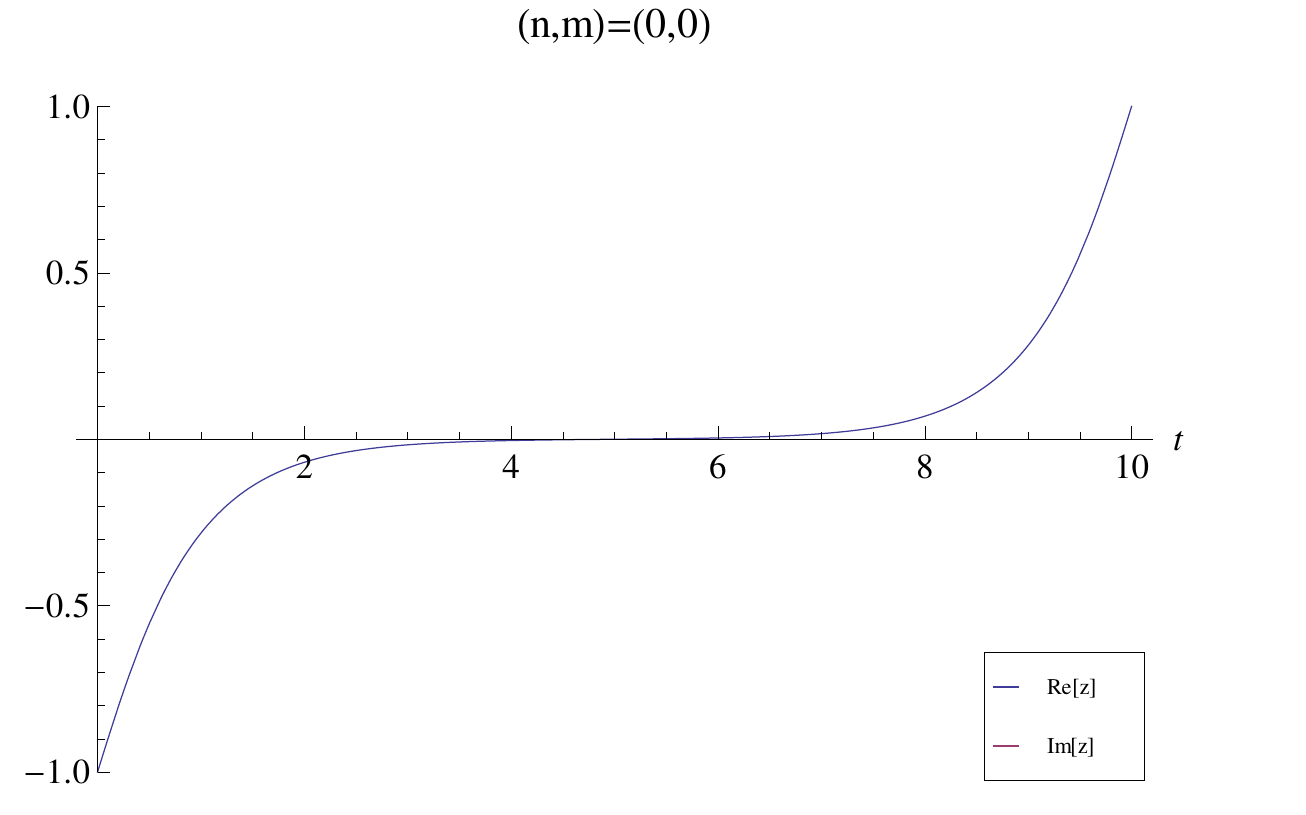}\;
\includegraphics[scale=0.38]{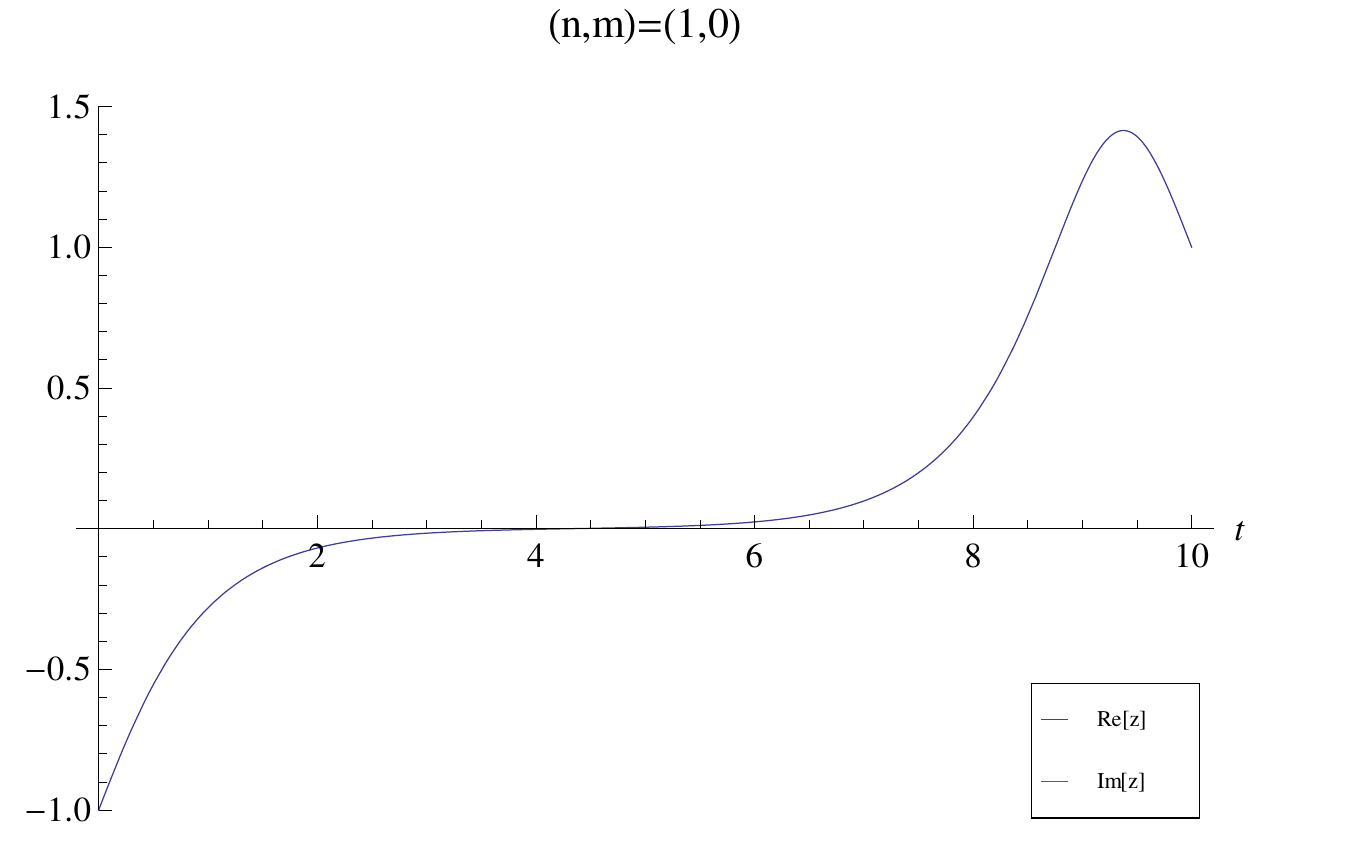}\;
\includegraphics[scale=0.38]{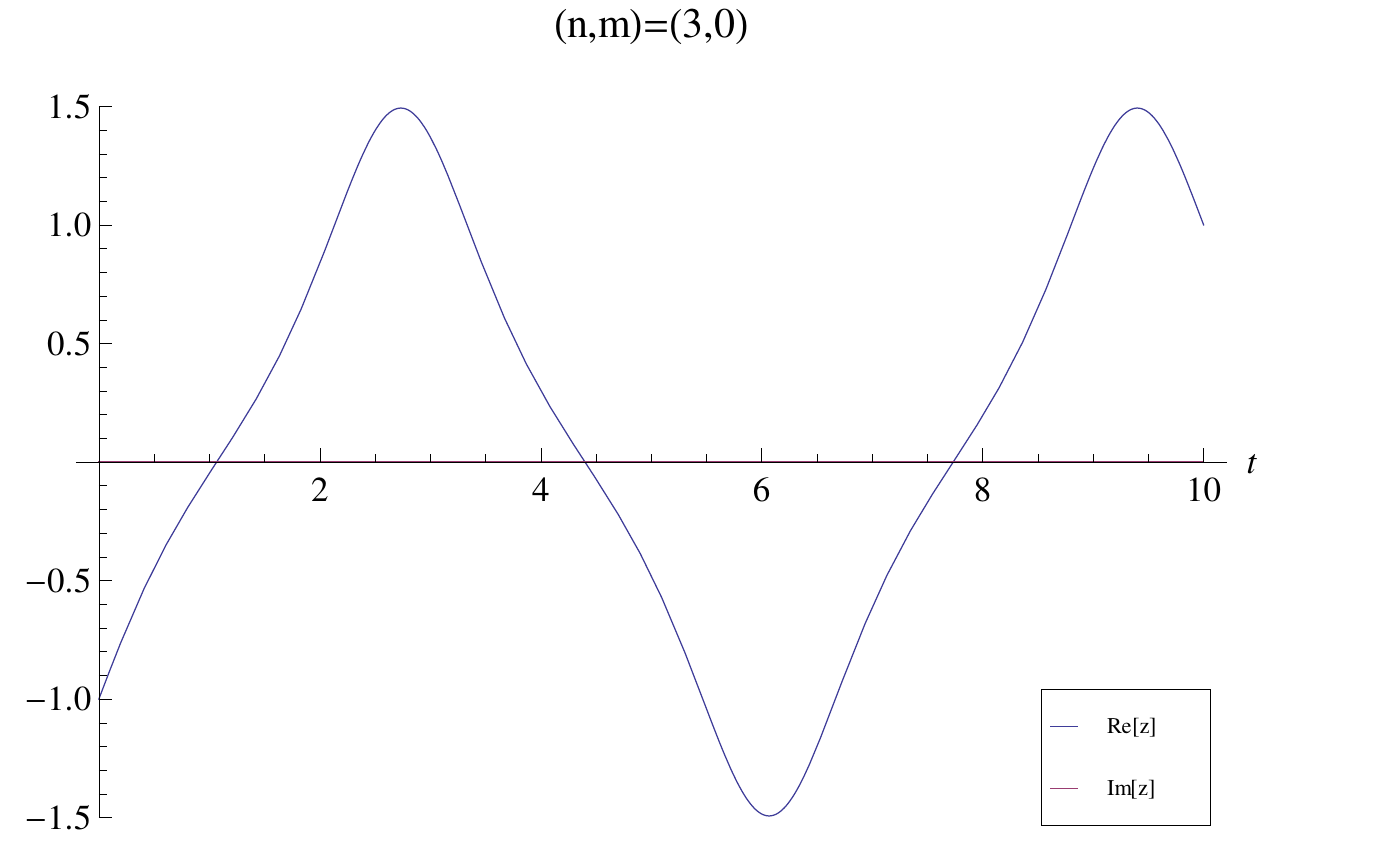}(b)
\caption{(a) Instanton-like solutions in the imaginary-time formalism. They correspond to $(n,m)=(0,0)$, $(1,0)$, and $(3,0)$, respectively, for $x_i=-1$, $x_f=1$, and $t_f-t_i=-10\im$ in (\ref{eq:relation_classical_solutions_integers02}). 
(b) Corresponding solutions in the real-time formalism. Especially, the solution with $(n,m)=(0,0)$ is very close to the unstable classical solution, and it seems to be natural to interpret them as sphalerons. }
\label{fig:instanton_and_sphaleron}
\end{figure}

Our purpose is to show a possibility to observe symmetry restoration within semiclassical analysis. For that purpose, we need to find a classical solution connecting two classical minima $x=\pm 1$ with small energies $p^2$. 
We will show that such behaviors can be realized only by complex classical solutions $z_{\sigma}$ with $\mathrm{Re}\;\mathcal{I}[z_{\sigma}]<0$. 

One possibility to find a real-time tunneling process is to consider an analytic continuation of instantons in imaginary-time formalism, but it turns out not to work. 
Let us try to consider it, however, because such consideration still gives an interesting suggestion: they are related to sphaleron processes. 
In Fig.~\ref{fig:instanton_and_sphaleron}~(a), instanton-like solutions in the imaginary-time formalism are shown for the case $x_i=-1$, $x_f=1$, and $t_f-t_i=-10\im$. 
Both of the first two ones ($(n,m)=(0,0)$ and $(1,0)$, respectively)  have almost the same value of the action $\mathcal{I}_0=-4/3$.
This value of the action is nothing but that of the one-instanton process. 
The last one $(n,m)=(3,0)$ has an action close to $3\mathcal{I}_0=-4$ and its deviation is quite small and comes from the finite-time effect \footnote{We got $\mathcal{I}_{(3,0)}\simeq-3.93 - 2.03\times 10^{-11} \im$. Resurgent trans-series theory \cite{Unsal:2012zj, Basar:2013eka, Cherman:2014ofa,Bogomolny1980431,Zinn-Justin1981125, ZinnJustin1983333, Dunne:2012ae, Dunne:2014bca} might be closely related to the smallness of $\mathrm{Im}(\mathcal{I}_{(n,0)})$. }, and thus it seems to be naturally interpreted as an instanton--anti-instanton--instanton process. 
In Fig.~\ref{fig:instanton_and_sphaleron}~(b), we show corresponding classical solutions in the real-time formalism. However, those processes are just real solutions of the classical equation of motion, and they do not suffer from exponential suppressions in the semi-classical limit $\hbar\to 0$. 
Rather, they seem to be related to unstable classical solutions, called sphalerons, in the long time limit $t_f-t_i\to \infty$, and then it is understandable why they are not suppressed in the semi-classical limit \cite{PhysRevD.30.2212, PhysRevD.36.581, PhysRevD.37.1020}. 
These processes are important to describe transitions among different topological sectors in quantum chromodynamics near the pseudo-critical temperature, but they are high-frequency phenomena and need sufficient energies, $p^2/2\simeq 1/2$, to overcome the potential barrier classically. 

In order to understand this behavior from an analytic viewpoint, let us consider about imaginary-time instantons in Fig.~\ref{fig:instanton_and_sphaleron}~(a) in more detail. 
An important property of those  solutions is that they have a quite small energy compared with the potential barrier: $|p^2|\ll 1$. 
Let us set the boundary condition $x_f=-x_i=1$ as above, and consider the limit $T\to \infty$ with $t_f=-t_i=-\im T/2$. 
If we set $(n,m)=(0,0)$ as an example, the boundary condition (\ref{eq:relation_classical_solutions_integers02}) becomes 
\be
-{\im\over 2}T=\sqrt{2k^2-1\over 2}\mathrm{sd}^{-1}\left(\sqrt{2k^2-1\over 2k^2(1-k^2)},k\right)
=-\im \sqrt{1\over 1+p}\mathrm{sn}^{-1}\left(\sqrt{1\over 1-p},\sqrt{1-p\over 1+p}\right). 
\label{eq:instanton01}
\ee
We rewrite the right hand side into a suitable form to study an asymptotic behavior in $p\to 0$ from the side $\mathrm{Im}\; p >0$ by using the formula given in Ref.~\cite[Sec.22.17]{NIST:DLMF}. Indeed, one can find in this limit that 
\be
\sqrt{1\over 1+p}\mathrm{sn}^{-1}\left(\sqrt{1\over 1-p},\sqrt{1-p\over 1+p}\right)\simeq {1\over 2}\ln {8\over p}+{\pi\over 4}\im. 
\label{eq:instanton02}
\ee
In the limit $T\to +\infty$, $p_{(0,0)}$ vanishes, and the well-known one-instanton solution can be obtained ($t=-\im \tau$):  
\be
p_{(0,0)}\simeq 8\im \exp-T,\quad z_{(0,0)}(\tau)=\sqrt{1-p_{(0,0)}}\mathrm{sn}\left(\sqrt{1+p_{(0,0)}}\tau,\sqrt{1-p_{(0,0)}\over 1+p_{(0,0)}}\right)\simeq \tanh \tau. 
\label{eq:instanton03}
\ee
We can study other instanton-like solutions with large imaginary times in the same way, and let us put $(n,m)=(1,0)$ for its demonstration. Then, the boundary condition (\ref{eq:relation_classical_solutions_integers02}) becomes 
\be
-{\im \over 2}T=\sqrt{2k^2-1\over 2}K(k)=\sqrt{1\over 1+p}\left\{K\left(\sqrt{2p\over 1+p}\right)-\im K\left(\sqrt{1-p\over 1+p}\right)\right\}. 
\label{eq:instanton04}
\ee
The right hand side behaves in the limit $p\to 0$ as $\pi/2-\im/2\ln(8/p)$, and then $p_{(1,0)}\simeq -8e^{-T}$, which again vanishes exponentially fast at $T\to \infty$. The corresponding solution $z_{(1,0)}$ is given by (again we set $t=-\im \tau$)
\bea
z_{(1,0)}(\tau)&=&\sqrt{1-p_{(1,0)}}\mathrm{sn}\left(\sqrt{1+p_{(1,0)}}(\tau+T/2)-\mathrm{sn}^{-1}\left(\sqrt{1\over 1-p_{(1,0)}},\sqrt{1-p_{(1,0)}\over 1+p_{(1,0)}}\right),\sqrt{1-p_{(1,0)}\over1+p_{(1,0)}}\right)\nonumber\\
&\simeq&\sqrt{1-p_{(1,0)}}\mathrm{sn}\left(\sqrt{1+p_{(1,0)}}\left(\tau+{T\over 2}-{1\over 2}\ln {8\over p_{(1,0)}}-{\pi\over 4}\im\right),{\sqrt{1-p_{(1,0)}\over 1+p_{(1,0)}}}\right)\nonumber\\
&\simeq&\tanh \left(\tau+{\pi\over 4}\im\right). 
\label{eq:instanton05}
\eea
This clarifies how the imaginary part appears in the solution $(n,m)=(1,0)$ of Fig.\ref{fig:instanton_and_sphaleron}~(a), and why its action has the same value of $\mathcal{I}_{0}=-4/3$ in the limit $T\to\infty$. 
According to these analyses, we can conclude that the singularity of (\ref{eq:relation_classical_solutions_integers02}) at $|k|\to \infty$, or $|p|\to 0$, plays an important role for instanton-like solutions in large imaginary times $(t_f-t_i)\to -\im \infty$. 
However, this singularity of the form ``$\im \ln p$'' cannot produce a large real part in (\ref{eq:relation_classical_solutions_integers02}), and real-time solutions satisfying $|p|\ll 1$ cannot be realized with fixed $(n,m)$ when $(t_f-t_i)\to+\infty$. This statement is true not only for real solutions with $(n,0)$ but also for general complex solutions with $(n,m)$. 

\begin{figure}[t]
\centering
\includegraphics[scale=0.25]{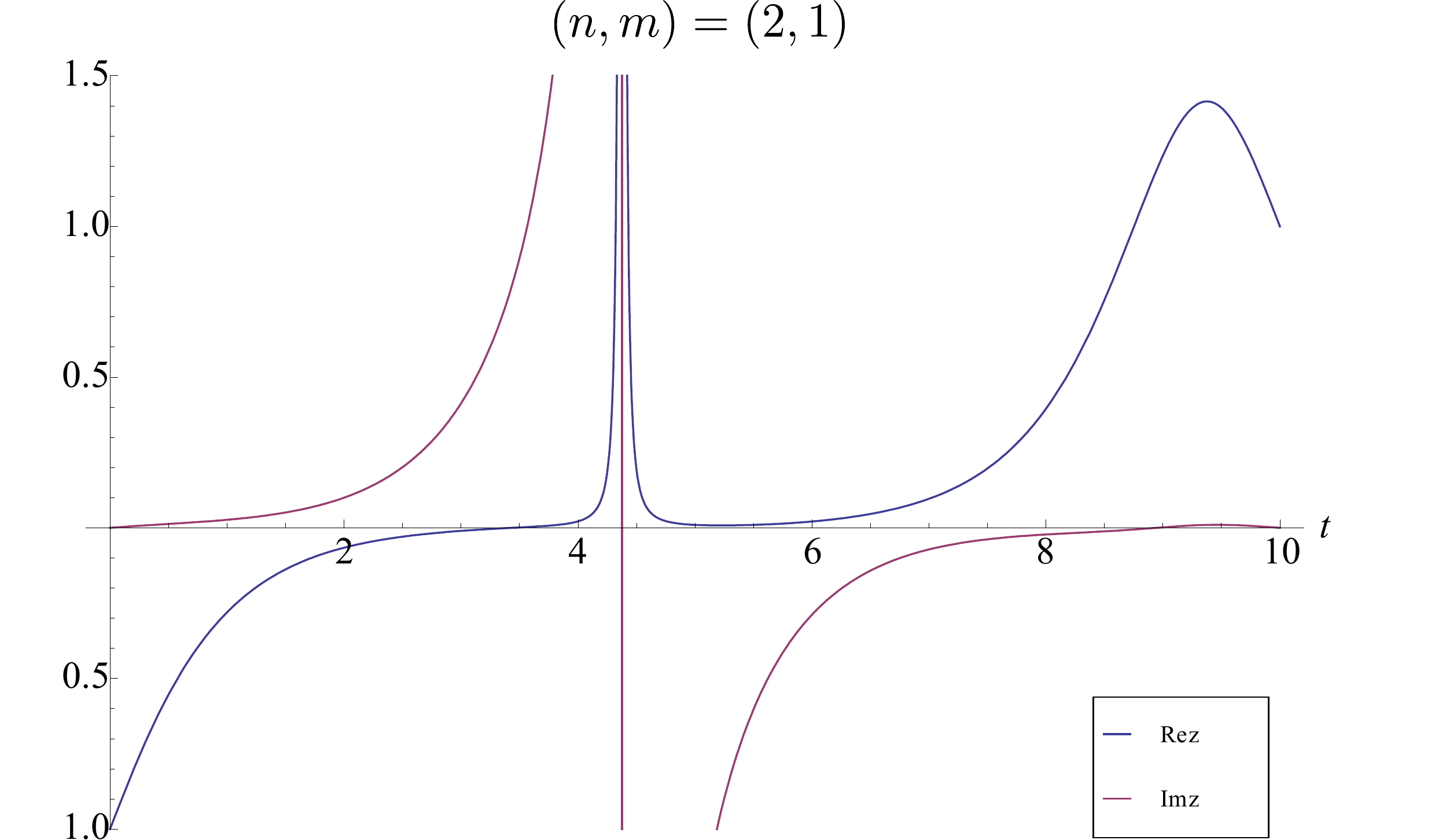}\quad
\includegraphics[scale=0.25]{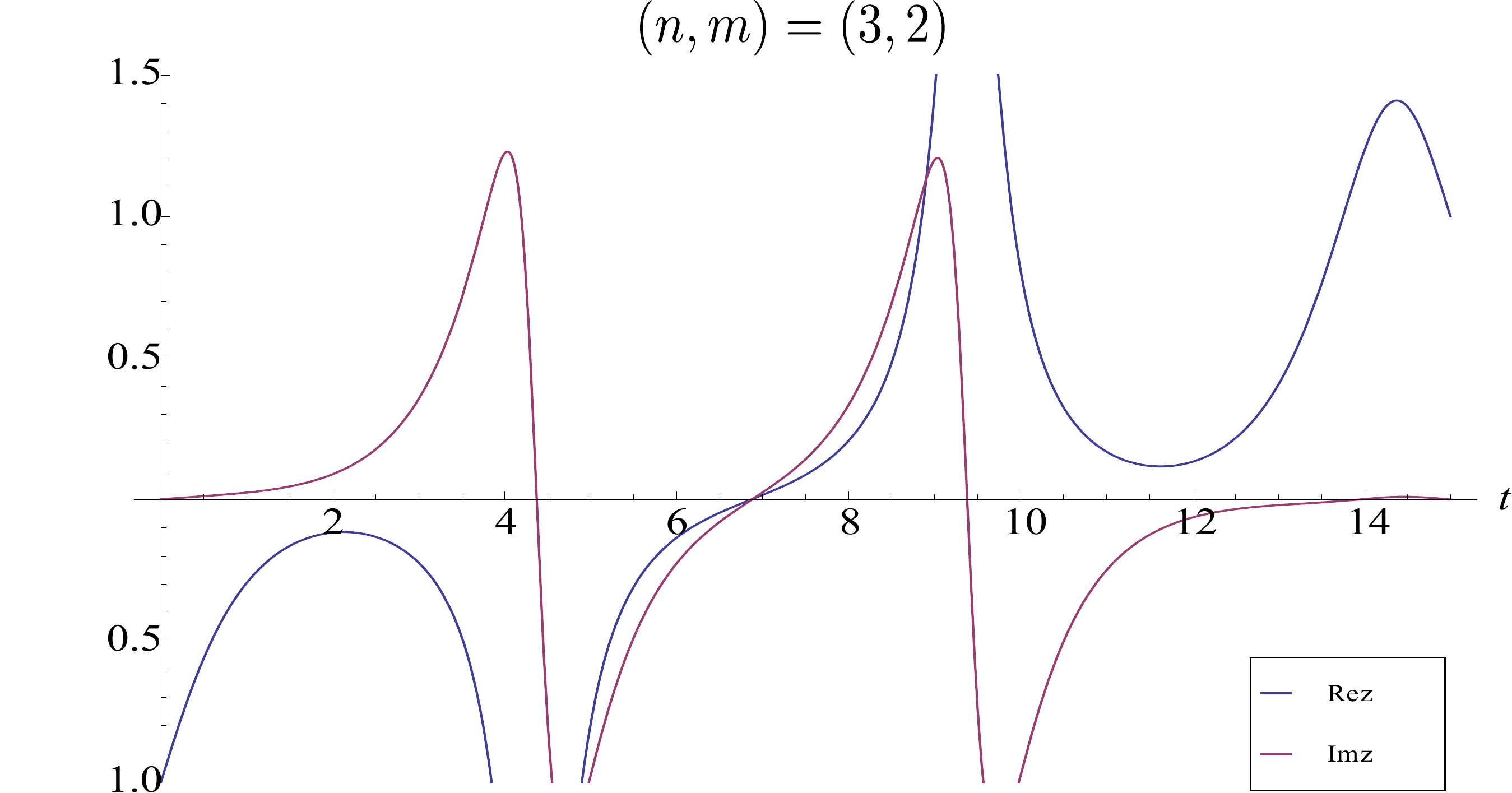}
\caption{Complex classical solutions $(n,m)=(2,1)$ with $(t_f-t_i)=10$ and $(n,m)=(3,2)$ with $(t_f-t_i)=15$. Complex parameters $p$ of these solutions are given by $1.001 + 0.027 \im$ and $0.987 + 0.024 \im$, respectively. They are close to the energy of the sphaleron process $p=1$. Actions of these solutions are given by $\mathcal{I}_{(2,1)}=-0.038-1.22\im$ and $\mathcal{I}_{(3,2)}=-0.051-1.871\im$, respectively. }
\label{fig:complex_sphalerons}
\end{figure}

Instead, $p$ converges to $\pm 1$ for fixed $(n,m)$ in large real times $(t_f-t_i)\to +\infty$, and we can also scrutinize asymptotic behaviors of real-time solutions in a similar way. Assume $n=m+1$ just for simplicity. 
Asymptotic analysis of (\ref{eq:relation_classical_solutions_integers02}) around $p=1$ gives 
\be
p_{(m+1,m)}\simeq 1+32\exp\left({m\over m+1}\pi \im-{\sqrt{2}(t_f-t_i)\over m+1}\right),\quad
\omega_1\simeq {t_f-t_i\over 2(m+1)}-{m\over m+1}{\pi\over 2\sqrt{2}}\im,\quad 
\omega_3\simeq {\pi\over 2\sqrt{2}}\im. 
\label{eq:instanton06}
\ee
By using the boundary condition $x(t_f)=-x(t_i)=1$ with $p\simeq 1$ and $(t_f-t_i)= 2\omega_1+2m(\omega_1+\omega_3)$, we get 
\be
z_{(m+1,m)}(t)\simeq \sqrt{p_{(m+1,m)}^2-1\over 2p_{(m+1,m)}}\mathrm{sd}\left(\sqrt{2p_{(m+1,m)}}\left(t-t_i-\omega_1+{\cosh^{-1}\sqrt{2}\over \sqrt{2}}\right),\sqrt{1+p_{(m+1,m)}\over 2p_{(m+1,m)}}\right). 
\label{eq:instanton07}
\ee
In Fig.~\ref{fig:complex_sphalerons}, we show complex solutions with $(n,m)=(2,1)$ and $(3,2)$ for $(t_f-t_i)=10$, and their typical behaviors can be explained by the approximated expression (\ref{eq:instanton07}). 
Half periods of this function is given by $2\omega_1$ and $2\omega_3$, thus the label $m$ refers the number of half periodicities of those solutions during the time $(t_f-t_i)\simeq 2\omega_1+m(2\omega_1+2\omega_3)$. 
Since the Jacobian elliptic function ``$\mathrm{sd}$'' has a pole at $(\omega_1+\omega_3)/\sqrt{2p}$ up to half periods, the approximated expression~(\ref{eq:instanton07}) diverges at some time if $\omega_3$ is real up to $2\omega_1$ and $2\omega_3$: 
This divergence occurs iff $m$ is an odd integer. 
The genuine solution does not diverge because of exponentially small corrections from (\ref{eq:instanton07}), and it explains why the solution $(2,1)$ has a quite sharp peak at $t-t_i\simeq (t_f-t_i)/2-(\cosh^{-1}\sqrt{2})/\sqrt{2}$ although such behavior cannot be observed in the solution $(3,2)$. 
We can also compute $\mathcal{I}_{(m+1,m)}$ in this limit $(t_f-t_i)\to +\infty$, and find that 
\be
\mathcal{I}[z_{(m+1,m)}]\simeq \im\left[-{1\over 2}(t_f-t_i)+{4\sqrt{2}\over 3}(m+1)\right]. 
\label{eq:instanton08}
\ee
Therefore, negative real parts of these actions $\mathcal{I}$ become smaller and smaller for larger real times $(t_f-t_i)$, and vanish as $(t_f-t_i)\to \infty$.  This is consistent with the smallness of $\mathrm{Re}(\mathcal{I})$ of solutions in Fig.~\ref{fig:complex_sphalerons}. 

\begin{figure}[t]\centering
\includegraphics[scale=0.3]{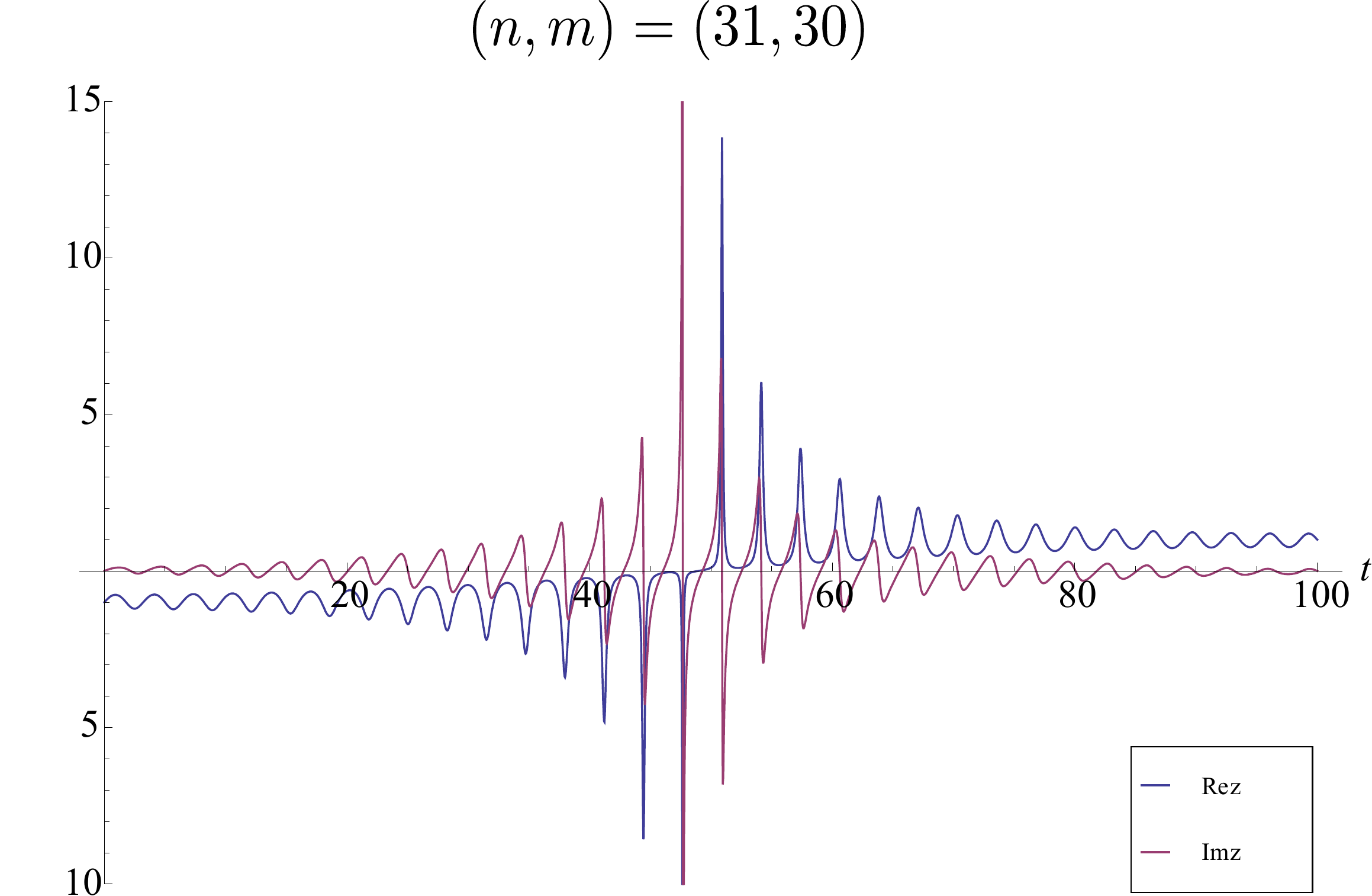}\quad 
\includegraphics[scale=0.32]{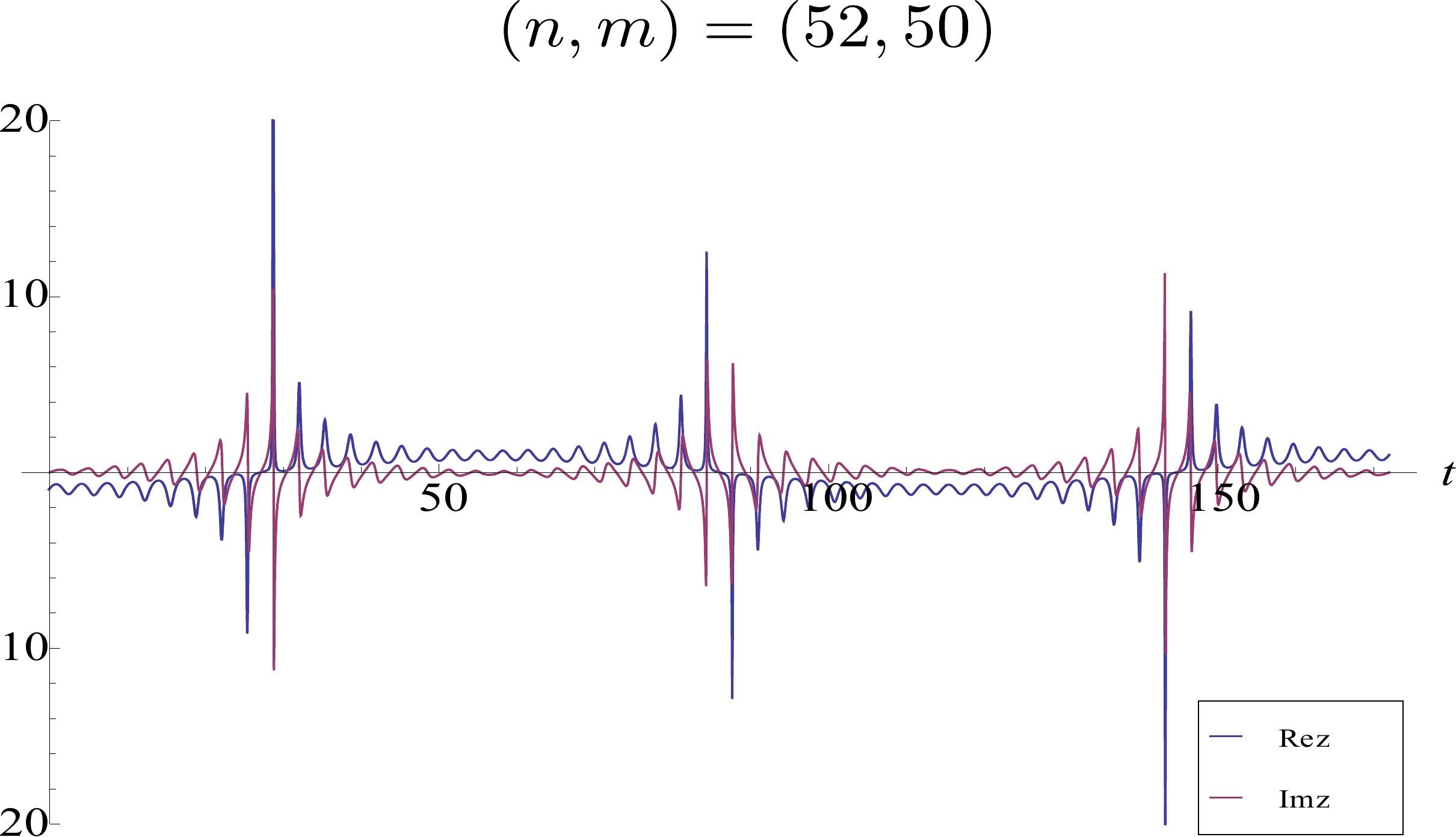}
\caption{Highly-oscillatory complex solutions. In spite of their oscillatory nature, actions of these solutions can be comparable with multi-instanton actions. Indeed, we obtained $p_{(31,30)}=0.427 + 0.155 \im$ and $\mathcal{I}_{(31,30)}=-1.072 + 0.007\im$ for $(n,m)=(31,30)$ with $(t_f-t_i)=100$, and $p_{(52,50)}=0.528+0.185\im$ and $\mathcal{I}_{(52,50)}=-2.892+0.092\im$ for $(n,m)=(52,50)$ with $(t_f-t_i)=172$, in these examples.  }
\label{fig:real_tunneling}
\end{figure}

The only possibility to describe the quantum tunneling in our formulation is the use of complex solutions with large labels $(n,m)$. 
In order to circumvent the constraint (\ref{eq:instanton08}) on the real-time complex solutions, the labels $n$, $m$ have to be at least of the order of $(t_f-t_i)$. 
Such solutions must be highly oscillatory, but they can have finite actions with non-vanishing negative real parts since they oscillate in the complexified configuration space (see Ref.~\cite{Cherman:2014sba} for detailed discussion on this property). 
Some examples are shown in Fig.~\ref{fig:real_tunneling}. 
Indeed, if $n_{\sigma}$ of such solutions are nonzero, then their classical actions have negative real parts  and those transition amplitudes are exponentially suppressed in the limit $\hbar\to 0$. 
Therefore, we must conclude that infinitely many complex solutions with large $(n,m)$ are significant in our path integral for describing the quantum oscillation. 

If the Picard--Lefschetz technique works well for the real-time description of quantum tunneling, infinitely many complex saddles must contribute, and intersection numbers $n_{\sigma}$ of those complex solutions $z_{\sigma}$ must be non-zero. 
Although it goes beyond our current abilities to compute intersection numbers $n_{\sigma}$, its importance is highlighted because of its close connection to physical phenomena, and we need more sophisticated understanding on downward/upward flow equations. 

\section{Summary}\label{sec:summary}
We studied  real-time Feynman path integrals of quantum mechanics from the viewpoint of Picard--Lefschetz theory. 
Quantum equation of motion is proven to hold on each Lefschetz thimble, and basic properties of downward/upward flows for the real-time formalism are discussed. 
In order to see how this method works in a concrete way, three simple examples of quantum mechanics are considered using path integrals on Lefschetz thimbles. Especially for the case of harmonic oscillator, appearance of Maslov--Morse index becomes quite clear by taking a close look on relative orientations between the Lefschetz thimble and the original cycle of the path integral. 

As a nontrivial example, quantum mechanics of a double-well potential is considered. We calculated all the complex saddle points $z_{\sigma}$ of the classical action, so we can get an enough data for semi-classical analysis on each Lefschetz thimble by computing Gaussian fluctuations around it. 
However, in order to relate those computations with the original path integral, we need to know how many of upward flows connect each saddle point to real spacetime paths. 
This gives an integral coefficient $n_{\sigma}$ of the path integral on Lefschetz thimble $\mathcal{J}_{\sigma}$. 
Some of them are determined by using general constraints on the flow equation, but there still exist infinitely many complex saddle points, whose intersection numbers $n_{\sigma}$ are not determined automatically. 

This problem occurs in general cases except for quadratic Lagrangians. 
If the Lagrangian has a non-quadratic term, then the number of complex saddle points becomes infinite in the continuum limit. 
Some of them will be complex and not real, which means that half of them have undetermined coefficients $n_{\sigma}$ and require us to analyze global structure of upward flows. 
However, upward flows are determined by nonlinear partial differential equations, and their global structure is not yet known. 
This is an open problem, which is very important for practical applications of this formalism to sign problems. 

In spite of this difficulty, we succeeded to derive a nontrivial consequence on quantum tunneling in real-time path integrals. 
In quantum mechanics, quantum tunneling restores symmetry and solves degeneracies between classical minima of the potential, so it is really nonperturbative phenomena. 
We first discuss analytic continuations of imaginary-time instantons, but they are related to sphaleron processes instead of real-time instantons. 
Since sphaleron processes have sufficient energies to overcome potential barriers, their amplitudes are free from exponential suppression in the semi-classical limit $\hbar\to 0$. 
The only possible description of real-time tunneling in our formalism is the use of highly oscillatory complex solutions, and their actions $\mathcal{I}[z_{\sigma}]$ indeed have non-vanishing negative real parts. 
This suggests that infinitely many complex saddles significantly contribute to real-time description of quantum tunneling. 
This statement is consistent with the fact that the tunneling amplitude behaves as $\exp-S_0/\hbar$ with some positive constant $S_0$, but we need to check $n_{\sigma}\not=0$ for such solutions in order to prove it. 
Although this goes beyond our current abilities, we believe that a new and solid understanding is now obtained for quantum tunneling based on exact semi-classical treatment of real-time path integrals. 

\section*{Acknowledgements}
Y.~T. thanks Takuya Kanazawa, Yoshimasa Hidaka, and Tetsuo Hatsuda for useful advices. He especially thanks Tetsuo Hatsuda also for carefully reading a manuscript of this paper. 
This work is completed when Y.~T. starts his half-year visit to University of Illinois at Chicago (UIC) and first presented in HEP/HEN seminar there, and he appreciates many comments in the seminar and hospitality by members of the theoretical nuclear and particle physics group at UIC, Mikhail~Stephanov, Ho-Ung~Yee, Wai-Yee~Keung, David~Mesterhazy, and Ahmed~Ismail.  
In the second manuscript of this paper, the authors added a detailed discussion on properties of the complex classical solutions for real-time quantum tunneling in Sec.~\ref{sec:discussion_tunneling} motivated by Ref.~\cite{Cherman:2014sba}, and Y.~T. appreciates private communication with its authors, Aleksey Cherman and Mithat {\"U}nsal. 
The authors, Y.~T. and T.~K., are supported by Grants-in-Aid for JSPS fellows (No.25-6615 and No.25-2869, respectively). 
This work is partially supported by  the RIKEN iTHES project, and also by the Program for Leading Graduate Schools, MEXT, Japan.

\appendix
\section{Short summary of Picard--Lefschetz theory}\label{sec:form}
In this appendix, the use of Picard--Lefschetz theory for oscillatory integrals is summarized for convenience of readers~\cite{Witten:2010cx,Witten:2010zr}.

Let $\mathcal{Y}$ be a real affine variety with a natural volume form $\diff \theta$. $S:\mathcal{Y}\to \mathbb{R}$ is a real polynomial on $\mathcal{Y}$, and we consider the oscillatory integral 
\be
Z_{\hbar}=\int_{\mathcal{Y}}\diff \theta \exp\left(\im S/\hbar\right), \label{osc_int_01}
\ee
with a real parameter $\hbar$. In order to reveal analytic properties on the $\hbar$-dependence of $Z_{\hbar}$, complex analysis often plays a crucial role. Therefore, we would like to analytically continue $Z_{\hbar}$ for generic complex parameters $\hbar$. 
The oscillatory integral (\ref{osc_int_01}), however, hinders its analytic property due to bad convergence. 
Picard--Lefschetz theory provides a beautiful framework, which converts such oscillatory integration into a sum of integrations with exponentially fast convergence. 

\subsection{Gradient flow and relative homology}
We assume the existence of complexification $\mathcal{X}$ of $\mathcal{Y}$, that is, $\mathcal{Y}$ is embedded in $\mathcal{X}$ and $\mathcal{Y}$ is fixed under the real involution operation $\bar{\cdot}$ of $\mathcal{X}$: $\mathcal{Y}\hookrightarrow \mathcal{X}$ and $y=\overline{y}$ for any $y\in \mathcal{Y}$. 
We introduce a \ka metric 
\be\diff s^2={1\over 2}g_{i\overline{j}}\left(\diff z^i\otimes \diff \overline{z^j}+\diff \overline{z^j}\otimes \diff z^i\right)
\ee 
on $\mathcal{X}$, and denote its \ka form as $\omega={\im\over 2}g_{i\overline{j}}\diff z^i\wedge \diff \overline{z^j}$. Here, $(z^1,\ldots,z^n)$ is a local holomorphic coordinate on $\mathcal{X}$. 

Regarding $\mathcal{I}=\im S/\hbar$ as a holomorphic function on $\mathcal{X}$, the Morse function $h$ is defined by its real part: 
\be 
h=\mathrm{Re}\; \mathcal{I}={\mathcal{I}+\overline{\mathcal{I}}\over 2}.
\ee
A point $p\in \mathcal{X}$ is a critical point of $h$ if and only if it is a critical point of $\mathcal{I}$. To see this, let $(z^1,\ldots,z^n)$ be a holomorphic local coordinate around $p$. Since $\mathcal{I}$ is holomorphic, the Cauchy--Riemann condition says $\overline{\p}_{\overline{i}}\mathcal{I}=0$. Therefore,  
\be
\p_i h=\overline{\p}_{\overline{i}}h=0\; \Leftrightarrow\; \p_i \mathcal{I}=0. 
\ee
Complex version of the Morse lemma suggests that there exists a local coordinate around a non--degenerate critical point $p$ such that 
\be
\mathcal{I}(z)=\mathcal{I}(0)+(z^1)^2+\cdots+(z^n)^2. 
\ee
Taking its real part, the Morse function behaves around the non--degenerate critical point as 
\be
h(z)=h(0)+(x^1)^2+\cdots+(x^n)^2-(y^1)^2-\cdots-(y^n)^2, 
\ee
with $z^j=x^j+\im y^j$. Therefore, index of the non--degenerate critical point is always $n=\mathrm{dim}_{\mathbb{C}}\mathcal{X}$. 

The downward flow equation is given by 
\be
{\diff z^i\over\diff t}=-2g^{i\overline{j}}{\p h\over \p \overline{z^j}}=-g^{i\overline{j}}\overline{\p}_{\overline{j}} \overline{\mathcal{I}}, \quad 
{\diff \overline{z^j}\over \diff t}=-2g^{i\overline{j}}{\p h\over \p z^i}=-g^{i\overline{j}} \p_i \mathcal{I}. 
\ee
This nomenclature is named after the fact that $h$ decreases monotonically along the flow: 
\be
{\diff h\over \diff t}=-2 g^{i\overline{j}}\p_i h \overline{\p}_{\overline{j}}h\le 0. 
\ee
The equality holds only at a critical point of $h$ on $\mathcal{X}$. There always exists a conserved quantity along the flow; $H=\mathrm{Im}\;\mathcal{I}=(\mathcal{I}-\overline{\mathcal{I}})/2\im$. Indeed, 
\be
{\diff H\over \diff t}={\diff z^i\over \diff t}\p_i H+{\diff \overline{z^j}\over \diff t}\overline{\p}_{\overline{j}}H=-{g^{i\overline{j}}\over 2\im}(\p_i \mathcal{I}\overline{\p}_{\overline{j}}\mathcal{I}+\p_i\mathcal{I}(-\overline{\p}_{\overline{j}}\mathcal{I}))=0. 
\ee
This phenomenon can be interpreted from the viewpoint of the classical mechanics~\cite{Witten:2010cx}. For that purpose, let us focus on the symplectic structure $\omega$ of $\mathcal{X}$, which defines the Poisson bracket $\{\cdot,\cdot\}_P$ by 
\be
\{f,g\}_P=-2\im g^{k\overline{l}}\left(\p_k f\overline{\p}_{\overline{l}}g-\overline{\p}_{\overline{l}}f\p_k g\right). 
\ee
Then, the downward flow equation can be written in the form of the Hamiltonian equation with the Hamiltonian $H=\mathrm{Im}\;\mathcal{I}$: 
\be
{\diff z^i\over \diff t}=\{H,z^i\}_P,\quad {\diff \overline{z^j}\over \diff t}=\{H,\overline{z^j}\}_P. 
\ee
This makes physically clear why $H=\mathrm{Im}\;\mathcal{I}$ is a conserved quantity along the flow. 

Let $\Sigma$ be the set of the labels $\sigma$ of the critical points, and the corresponding critical point is denoted as $p_{\sigma}$, i.e., $\p\mathcal{I}|_{p_{\sigma}}=0$ for any $\sigma\in\Sigma$.  
We would like to find out all possible integration cycles for the integration form $\exp(\im S/\hbar)\diff \theta$, and then the integrand must decrease sufficiently fast at infinities. This indicates that all the possible integration cycles can be identified as an element of the relative homology $H_n(\mathcal{X},\mathcal{X}_{-T};\mathbb{Z})$ for $T\gg 1$, where 
\be
\mathcal{X}_{-T}:=\{x\in X\;|\; h(x)\le -T\}. 
\ee
Therefore, it is of great importance to identify the relative homology $H_n(\mathcal{X},\mathcal{X}_{-T};\mathbb{Z})$, and its generators are called Lefschetz thimbles. 

Let $\sigma\in \Sigma$, and consider the downward flows starting from $p_{\sigma}\in \mathcal{X}$. Assume that any downward flows does not connect distinct critical points, then the Morse function diverges to $-\infty$ as the flow time goes to $\infty$. We define the Lefschetz thimble $\mathcal{J}_{\sigma}$ associated to the critical point $p_{\sigma}$ as the moduli space for endpoints of solutions $c:\mathbb{R}\to \mathcal{X}$ of the downward flow equation 
\be
{\diff c \over \diff t }=\{\mathrm{Im}\;\mathcal{I},c\}_P, 
\ee
with the initial condition $c(-\infty)=p_{\sigma}$. That is, 
\be
\mathcal{J}_{\sigma}=\left\{c(0)\in \mathcal{X}\;\left|\; \dot{c}(t)=\{\mathrm{Im}\;\mathcal{I},c(t)\}_P\;,\; c(-\infty)=p_{\sigma}\right.\right\}. 
\ee
Clearly, there exists $n$ independent directions for the downward flows, and thus the moduli space $\mathcal{J}_{\sigma}$ has real dimension $n$. 
Since $h(c(t))\to -\infty$ as $t\to \infty$ except for $c(t)=p_{\sigma}$, the Lefschetz thimble  $\mathcal{J}_{\sigma}$ defines an element of the relative homology $H_n(\mathcal{X},\mathcal{X}_{-T};\mathbb{Z})$. 
The important and very strong facts are that all the elements of $H_n(\mathcal{X},\mathcal{X}_{-T};\mathbb{Z})$ are generated by the Lefschetz thimbles associated to critical points and that other homologies vanish 
\footnote{These are because we here only treat, so called, harmonic Morse functions. For general Morse functions, we must compensate with a weaker statement. }. 

\subsection{Real cycle in terms of Lefschetz thimbles}
We would like to express the real cycle $\mathcal{Y}$ in terms of Lefschetz thimbles as 
\be
\mathcal{Y}=\sum_{\sigma\in \Sigma}n_{\sigma}\mathcal{J}_{\sigma}, 
\ee
where $n_{\sigma}\in\mathbb{Z}$. With this expression, the oscillatory integral becomes 
\be
Z_{\hbar}=\sum_{\sigma\in \Sigma} n_{\sigma}\int_{\mathcal{J}_{\sigma}}\diff \theta \exp(\im S/\hbar). 
\ee
Since $\mathrm{Im}(iS/\hbar)$ is constant on each Lefschetz thimble $\mathcal{J}_{\sigma}$, this expression does not contain oscillatory integration. 

For this purpose, we need to find some pairing operation. We define 
\be
\mathcal{X}^T=\{x\in \mathcal{X}\;|\; h(x)\ge T\}, 
\ee
and consider the relative homology $H_n(\mathcal{X},\mathcal{X}^T;\mathbb{Z})$. 
As in the case of $H_n(\mathcal{X},\mathcal{X}_{-T};\mathbb{Z})$ for $T\gg 1$, the relative homology $H_n(\mathcal{X},\mathcal{X}^T;\mathbb{Z})$ is generated by the upward flowing cycles 
\be
\mathcal{K}_{\sigma}=\left\{c(0)\in \mathcal{X}\;|\; \dot{c}(t)=-\{\mathrm{Im}\;\mathcal{I},c(t)\}_P\;,\; c(-\infty)=p_{\sigma}\right\}. 
\ee
Among these bases $\{\mathcal{J}_{\sigma}\}_{\sigma\in\Sigma}$ and $\{\mathcal{K}_{\tau}\}_{\tau\in\Sigma}$, the intersection pairing is naturally given as 
\be
\langle \mathcal{J}_{\sigma},\mathcal{K}_{\tau}\rangle=\delta_{\sigma\tau}
\ee
under some appropriate orientations. Therefore, the coefficients $n_{\sigma}$ can be calculated as 
\be
n_{\sigma}=\langle \mathcal{Y},\mathcal{K}_{\sigma}\rangle. 
\ee
That is, $n_{\sigma}$ counts the number of intersecting points between the real cycle $\mathcal{Y}$ and the upward flowing cycles emanating from the critical point $p_{\sigma}$. 

According to this fact, it is convenient to decompose the label set $\Sigma$ into three parts; $\Sigma=\Sigma_{\mathbb{R}}\oplus \Sigma_+\oplus\Sigma_-$, with 
\bea
\Sigma_{\mathbb{R}}&=&\{\sigma\in \Sigma \;|\; p_{\sigma}\in \mathcal{Y}\}, \\
\Sigma_+&=& \{\sigma \in \Sigma \;|\; p_{\sigma}\not\in \mathcal{Y},\; h(p_{\sigma})\ge 0  \}, \\
\Sigma_-&=& \{\sigma \in \Sigma \; | \; p_{\sigma}\not\in \mathcal{Y},\; h(p_{\sigma})<0\}. 
\eea
We should notice that $h(y)=\mathrm{Re}(\im S(y)/\hbar)=0$ for real $\hbar$ and $y\in \mathcal{Y}$. 
This implies the strong constraint on the coefficient $n_{\sigma}$ for $\sigma\in\Sigma_{\mathbb{R}}$ and $\sigma\in \Sigma_+$. 
Let $\tau \in \Sigma_{\mathbb{R}}$, and consider the upward flowing cycle $\mathcal{K}_{\tau}$. Then, $h(p_{\tau})=0$ since $p_{\tau}\in \mathcal{Y}$, but $h(p)>0$ for any $p\in \mathcal{K}_{\tau}\setminus \{p_{\tau}\}$ by definition of the upward flow. Therefore, 
\be
\langle \mathcal{Y},\mathcal{K}_{\tau}\rangle=1. 
\ee
On the other hand, $h(p)\ge h(p_{\tau})\ge 0$ for $\tau \in \Sigma_+$ and $p\in \mathcal{K}_{\tau}$, but $p_{\tau}\not\in \mathcal{Y}$. Thus, 
\be
\langle \mathcal{Y},\mathcal{K}_{\tau}\rangle =0
\ee
for $\tau \in \Sigma_+$. The oscillatory integral turns out to be written as 
\be
Z_{\hbar}=\sum_{\sigma\in \Sigma_{\mathbb{R}}}\int_{\mathcal{J}_{\sigma}}\diff \theta \exp(\im S/\hbar) +\sum_{\sigma\in\Sigma_-}n_{\sigma}\int_{\mathcal{J}_{\sigma}}\diff \theta \exp\left(\im S/\hbar\right). 
\label{app:PicardLefschetz}
\ee
The intersection numbers $n_{\sigma}$ for elements of $\Sigma_-$ cannot be determined only from general properties of flow equations, and we need to have a close look at behaviors of upward flows for each case.



\bibliographystyle{utphys}
\bibliography{./lefschetz,./QFT}


\end{document}